\documentclass[12pt]{JHEP3}
\pdfoutput=1

\usepackage{amsmath,epsfig}
\usepackage{amssymb,amsfonts}
\usepackage{latexsym}
\usepackage[latin1]{inputenc}
\usepackage{subeqnarray}
\usepackage{xcolor}
\let\normalcolor\relax
\usepackage{graphicx}

\usepackage{longtable}

\relax
\renewcommand{\theequation}{\arabic{section}.\arabic{equation}}
\def\be{\begin{equation}}
\def\ee{\end{equation}}
\def\bea{\begin{eqnarray}}
\def\eea{\end{eqnarray}}

\newcommand\fverb{\setbox\pippobox=\hbox\bgroup\verb}
\newcommand\fverbdo{\egroup\medskip\noindent%
                        \fbox{\unhbox\pippobox}\ }
\newcommand\fverbit{\egroup\item[\fbox{\unhbox\pippobox}]}

\newcommand{\bear}{\begin{eqnarray}}

\newcommand{\eear}{\end{eqnarray}}

\newcommand{\bsea}{\begin{subeqnarray}}
\newcommand{\esea}{\end{subeqnarray}}
\newbox\pippobox

\def\N{{\cal N}}

\def\6{\partial}
\def\f{\Phi}
\def\a{\alpha}
\def\g{\gamma}
\def\nn{\nonumber}

\def\e{\epsilon}

\def\n{\nu}

\def\sp{\;\;\;,\;\;\;}

\def\sq
\def\a{\alpha}

\def\y{\psi}

\def\k{\chi}

\def\e{\epsilon}

\def\Z{\Zeta}
\def\L{\Lambda}

\def\o{\over}

\title{\huge Backreacted DBI Magnetotransport with Momentum Dissipation}
\author{\large   Sera Cremonini,  Anthony Hoover and Li Li\\
~\\
Department of Physics, Lehigh University, Bethlehem, PA, 18018 USA.
~\\

E-mail: cremonini@lehigh.edu, arh314@lehigh.edu,  lil416@lehigh.edu}

\abstract{We examine magnetotransport in a holographic Dirac-Born-Infeld model,  
taking into account the effects of backreaction on the geometry. The theory we consider includes axionic scalars,
introduced to break translational symmetry and generate momentum dissipation. 
The generic structure of the DC conductivity matrix for these theories is extremely rich, 
and is significantly more complex than that obtained in the probe approximation.
We find new classes of black brane solutions, including geometries that exhibit Lifshitz scaling and hyperscaling violation, and examine their implications on the transport properties of the system. Depending on the choice of theory parameters, these backgrounds can 
lead to metallic or insulating behavior. Negative magnetoresistance is observed in a family of dynoic solutions.
Some of the new backreacted geometries also support
magnetic-field-induced metal-insulator transitions.
}

\keywords{AdS/CMT, Quantum Criticality, Magnetoresistance}

\def\pdfannotlink{\pdfstartlink}
\begin{document}

\newpage

\section{Introduction}
\label{intro}

Holographic techniques have provided new avenues for exploring the behavior of strongly coupled quantum phases of matter.
In recent years much of the focus has been on understanding the transport properties of models that may be in the same universality class of strongly correlated electron systems, whose unconventional behavior is believed to be tied to the richness and complexity of their phase diagram (see {\em e.g.}~\cite{Hartnoll:2016apf} for a recent review). A particularly puzzling behavior is that of the linear temperature dependence of the electrical resistivity $R \sim T$ displayed by many correlated electron metals, which is often associated with the existence of an underlying quantum critical point (QCP).
This strange metal behavior has been argued to be due to the fact that the temperature is the dominant energy scale in the system, and 
therefore sets the scattering rate near a QCP, resulting in the $T$-linear resistivity. 

A natural question is then whether the same argument applies to 
magneto-transport phenomena in quantum
critical systems, with the magnetic field behaving much like temperature.
Indeed, it has been shown recently~\cite{Hayers:2014} that in the pnictide superconductor BaFe$_2($As$_{1-x}$P$_x)_2$  near its QCP 
the magnetic field $h$ plays the same role as the temperature $T$, with 
the in-plane resistance described well by 
\begin{equation}\label{experiment}
R_{DC}=\sqrt{\hat{\alpha} \,T^2+\hat{\eta}\, h^2} \, ,
\end{equation}
where $\hat{\alpha},\hat{\eta}$ are constants which relate the scattering rate to the temperature and magnetic field scales (see~\cite{Hayers:2014} for more details).
It is believed that in quantum critical metals $h$ and $T$ compete with each other to set the scale of the scattering rate, and thus 
magnetic fields provide yet another way to probe the unconventional linear resistivity of the strange metal phase. 

It was shown in~\cite{Kiritsis:2016cpm} that the behavior (\ref{experiment}) can be generated holographically by working with a string-inspired Dirac-Born-Infeld (DBI) model, which can be thought of as a non-linear realization of 
electrodynamics which encodes the low-energy dynamics of D-branes. 
In particular, the result  (\ref{experiment}) is a special case of a broader class of allowed scaling behaviors, which are realized by considering 
finite temperature backgrounds exhibiting hyperscaling violation $\theta\neq 0$ and a non-trivial dynamical exponent $z\neq 1$. A specific choice of exponents $z$ and $\theta$ then yields precisely~\eqref{experiment}.

However, the analysis of~\cite{Kiritsis:2016cpm} was done in the probe approximation, building on the work of~\cite{Karch:2007pd,OBannon:2007cex}, 
in which the backreaction of the charge density and magnetic field on the geometry is neglected. 
Indeed, in the probe DBI limit 
the charge degrees of freedom are subleading as compared to the uncharged ones (the D-branes leave the background unchanged since their backreaction is not taken into account). 
As a result, the coefficient of the momentum conserving $\delta$-function is hierarchically suppressed and the DC conductivity is finite even though
the system has translational invariance.
However, once the full backreaction of the DBI action is taken into account, the above probe description breaks down and one recovers the usual infinite DC conductivity due to the presence of translational symmetry.

In this paper we are going to extend the results of~\cite{Kiritsis:2016cpm,Karch:2007pd,OBannon:2007cex} by going beyond the probe limit and examining the effect of backreaction of the DBI action on the geometry.
In order to introduce momentum dissipation and ensure that the theory can lead to a finite DC conductivity, we add axionic scalars~\cite{Andrade:2013gsa}, thus breaking translational invariance.
The axions are taken to depend on the spatial directions linearly, making the analysis tractable. 
We will find new classes of solutions, including geometries that exhibit Lifshitz scaling and hyperscaling violation, which can be associated with new quantum critical regimes.
Armed with these backgrounds, we will examine the implications on the transport properties of the dual system.
The question we are interested in is that of the role of the fully non-linear effects encoded by the DBI interactions on the conductive properties -- are there any features inherent to the backreacted analysis that would be absent in the probe approximation? 
Our focus will be on the interplay between temperature and magnetic field, in the presence of momentum relaxation.
We will find a very rich structure for the resistivity that arises in this class of models.



\subsection{Summary of Results}

We have examined the behavior of the DC conductivity/resistivity matrix as a function of the physical scales in the problem -- temperature, charge density, magnetic field and momentum dissipation\,\footnote{Due to a scaling symmetry, only three of these four scales are 
actually physical.} -- taking into account the full backreaction of the D-brane action on the geometry.
Axionic fields were used to break translational invariance and ensure momentum relaxation.
A dilatonic scalar appropriately coupled to the DBI interaction term was introduced to generate scaling solutions.
We have found a highly complex and rich structure for the magnetotrasport, given 
in expressions~\eqref{DC}~\eqref{resis}, which simplifies somewhat in a number of limiting cases.
In full generality, the dependence of the conductivity matrix on the physical scales in the system and the couplings of the theory is significantly more complicated than that of the probe DBI limit, which is summarized in~\eqref{strongkresult}. 

In general the various contributions to the DC conductivity combine in a non-trivial fashion 
-- the terms associated with momentum relaxation of the charge carriers in the system and those independent of the charge degrees of freedom 
are not added together in a simple way. Thus, this provides an explicit example in which there is no clean separation between coherent and incoherent contributions.
The results of the probe DBI approximation are recovered in the limit of strong momentum dissipation, for which the 
contribution of the DBI action to the geometry is negligible as compared to that of the axionic sector.
In the opposite limit of weak momentum relaxation, the conductivity tensor to leading order is independent of temperature and of the details of the theory, as a  consequence of Lorentz invariance \cite{Hartnoll:2007ai}. 
Only at next-to-leading order one finds non-trivial dependence on $T$ and the specific parameters of the model.

We have identified several classes of new, exact solutions to the theory and discussed the physical constraints on the parameter space needed to have a well-defined holographic ground-state.  
Depending on the theory parameters, these solutions can describe either metallic or insulating phases.
While they are valid everywhere in the geometry (they are exact), 
when their UV asymptotics are not AdS we will interpret them as describing only the IR of the geometry, in order to 
adopt the standard holographic AdS/CFT dictionary. That they can be embedded in AdS (by making minor modifications to the scalar potential) is by now well known.

When the dilatonic scalar is trivial, the exact black-brane geometries are associated with a metal-insulator crossover, induced by varying the magnitude of the magnetic field. On the other hand, a running dilatonic scalar leads to exact hyperscaling violating, Lifshitz-like black brane solutions, 
which also exhibit either metallic or insulating behavior, depending on the range of parameters.
For some of the simpler classes of scaling solutions we have obtained, we find that the DC conductivity 
in the absence of magnetic field scales with temperature as 
\begin{equation}
\label{scalingsigma}
\sigma_{DC} \sim T^{\frac{\theta-4}{z}}\,,
\end{equation}
yielding a linear resistivity $R_{DC}=1/\sigma_{DC} \sim T$ along the line $\theta+z=4$ which is allowed in much of the physical parameter space of the theory.  Interestingly, we have also identified a somewhat simple hyperscaling violating solution with non-vanishing magnetic field, 
and with $\theta=4$. For this solution the result~\eqref{scalingsigma} still applies and $\sigma_{DC}$ is constant.
Thus, this special $\theta=4$ geometry sits at the edge of the insulating 
and metallic behavior seen in~\eqref{scalingsigma}. 
Moreover, for this dyonic case we observe a negative magnetoresistance, a feature which is absent in the probe DBI limit.
Exact solutions with non-zero magnetic field and more arbitrary values of $\theta$ can also be identified, but are significantly more complicated. We expect them to lead to a similarly rich structure for the magnetotransport, and leave their analysis to future work. 

The key message to take away from our analysis is that by taking into account backreaction, the transport behavior which can be realized in this theory is rich and highly complex. Non-trivial classes of IR geometries can be easily constructed, which allow for a wide range of scalings. 
They give rise to not only metallic or insulating behavior, but also new magnetic field driven metal-insulator crossovers as well as a negative magnetoresistance.
In this paper we have only begun to explore the properties of these solutions, and their implications for transport. 
We anticipate that disorder driven transitions (driven by changing the magnitude $k$ of the axionic scalars) may also be 
possible to realize in these models, perhaps using new classes of black brane solutions.
It would also be interesting to construct the full geometries that interpolate between the IR solutions 
we have identified and the $AdS_4$ fixed point in the UV, and study their AC transport properties.
We leave the exploration of these questions to future studies. 
Before closing, we note that while we were in the last stages of this project, the paper~\cite{Blake:2017qgd} appeared, whose conductivity analysis for DBI Q-lattice models partially overlaps with our results. 

The outline of the paper is as follows. Section~\ref{setup} introduces our holographic DBI model while
Section \ref{ConductivitiesSection} contains the computation of the DC conductivity matrix and a discussion of simple limiting cases. In Section \ref{metalinsulator} we present exact black brane solutions for the case of a trivial dilatonic scalar, and discuss the associated magnetic-field-induced metal-insulator transition. Section \ref{SectionNonRelativistic} contains new exact hyperscaling violating and Lifshitz-like scaling geometries and discusses the associated transport behavior.
Finally, in Appendix \ref{Appendix} we include for complenetss the magnetotransport analysis for the simpler Born-Infeld theory.

\section{The Holographic Setup}\label{setup}
We consider a four-dimensional model describing gravity coupled to a neutral scalar field $\phi$, two axions $\psi^I$ and an abelian gauge field $A_\mu$, whose dynamics is described by the DBI action, 
\begin{equation}\label{action}
S=\int d^{4}x\sqrt{-g}\left[\mathcal{R}-{1\over 2}(\partial\phi)^2-V(\phi)-\frac{Y(\phi)}{2}\sum_{I=1}^2 (\partial \psi^I)^2  \right]+S_{DBI}\,,
\end{equation}
with\,\footnote{The second term in $S_{DBI}$, which could have been incorporated into $V(\phi)$, 
is chosen to make it apparent that in the weak flux limit $F \rightarrow 0$ one recovers the standard gauge 
field kinetic term.} the DBI term
\begin{equation}\label{DBI}
\begin{split}
S_{DBI}&=-\int d^{4}x Z_1(\phi)\left[\sqrt{-\det(g_{\mu\nu}+Z_2(\phi)F_{\mu\nu}})-\sqrt{-\det (g_{\mu\nu})}  \; \right].\\
\end{split}
\end{equation}
The couplings $Z_1(\phi), Z_2(\phi), Y(\phi)$ are introduced to lead to non-trivial interactions between the scalar sector and the gauge field. 
It is well known that dimensional reductions usually involve several matter fields and non-trivial potentials for the lower-dimensional scalars. 
It would be interesting to find an actual top-down construction in which the couplings of the D-brane action are fixed uniquely (see, for example,~\cite{Kiritsis:2016cpm}).
However in this paper we focus on a bottom-up approach and view (\ref{DBI}) as an effective theory, without worrying about its detailed string theory origin. In particular, we would like to see whether this effective theory can lead to interesting behaviors for the magnetotransport of the putative dual quantum system. 

The equations of motion associated with the action~\eqref{action} take the form
\begin{equation}\label{eomphi}
\begin{split}
\nabla_\mu\nabla^\mu\phi-V'(\phi)-\frac{Y'(\phi) }{2}\sum_{I=1}^2 (\partial \psi^I)^2-Z_1'(\phi)\left[\sqrt{\frac{-\det (g+Z_2(\phi) F)}{-\det g}}-1\right]\\
+\frac{Z_1(\phi) Z_2'(\phi)}{2}\sqrt{\frac{-\det (g+Z_2(\phi) F)}{-\det g}} (g+Z_2(\phi)F)^{-1[\mu\nu]}F_{\mu\nu}=0\,,
\end{split}
\end{equation}
\begin{equation}\label{eomA}
\nabla_\mu \left[Z_1(\phi)Z_2(\phi)\sqrt{\frac{-\det (g+Z_2(\phi) F)}{-\det g}}(g+Z_2(\phi)F)^{-1[\mu\nu]}\right]=0\,,
\end{equation}
\begin{equation}\label{eompsi}
\nabla_\mu\left(Y(\phi)\nabla^\mu \psi^I\right)=0\, ,
\end{equation}
%
%
\begin{equation}\label{eomg}
\begin{split}
\mathcal{R}_{\mu\nu}-{1\over 2}\mathcal{R}g_{\mu\nu}=\frac{1}{2}\left(\partial_\mu\phi\partial_\nu\phi-\frac{1}{2}g_{\mu\nu}(\partial\phi)^2\right)
+\frac{Y(\phi)}{2}\sum_{I=1}^2\left(\partial_\mu\psi^I\partial_\nu\psi^I-\frac{1}{2}g_{\mu\nu}(\partial\psi^I)^2\right)\\
-\frac{1}{2}g_{\mu\nu}V(\phi) +T^{DBI}_{\mu\nu}\,,
\end{split}
\end{equation}
with the DBI stress energy tensor  $T^{DBI}_{\mu\nu}=-\frac{1}{\sqrt{-g}}\frac{\delta S_{DBI}}{\delta g^{\mu\nu}}$ given by
\begin{equation}
T^{DBI}_{\mu\nu}=
-\frac{Z_1(\phi)}{2}\sqrt{\frac{-\det (g+Z_2(\phi) F)}{-\det g}}g_{\mu\alpha}(g+Z_2(\phi) F)^{-1(\alpha\beta)}g_{\beta\nu}+\frac{Z_1(\phi)}{2}g_{\mu\nu}\,.
\end{equation} 
Here $(g+Z_2(\phi)F)^{-1\mu\nu}$ is the inverse of $(g+Z_2(\phi)F)_{\mu\nu}$, with the subscript $^{(\;)}$ denoting the symmetric part (and $^{[\;]}$ the antisymmetric part).
The current in the dual field theory, evaluated at the boundary, reads
\begin{equation}\label{current}
\begin{split}
J^\mu&=\sqrt{-\gamma} \, n_\nu \,  Z_1(\phi)Z_2(\phi)\sqrt{\frac{-\det (g+Z_2(\phi) F)}{-\det g}}  \left(g+Z_2(\phi)F \right)^{-1[\nu\mu]} \Big{|}_\partial\\
&=Z_1(\phi)Z_2(\phi)\sqrt{-\det (g+Z_2(\phi) F)} \left(g+Z_2(\phi)F\right)^{-1[r\mu]} \Big{|}_\partial\,,
\end{split}
\end{equation}
The quantities $\gamma$ and $n^\mu$ in this expression are, respectively, the induced metric and outward pointing normal vector at the asymptotically AdS boundary. Here we have used $r$ to denote the holographic radial direction.

Assuming homogeneity and isotropy, the bulk metric and the matter fields take the generic form,
\begin{eqnarray}\label{fullansatz}
\begin{split}
ds^2=-D(r)dt^2+B(r)dr^2+C(r) (dx^2+dy^2),\quad \phi=\phi(r),\\
\psi^1=k\, x,\quad\psi^2=k\, y,\quad A=A_{t}(r)\,dt+\frac{h}{2}(x dy-y dx)\,,
\end{split}
\end{eqnarray}
with $h$ denoting the magnitude of the magnetic field.
The axions depend on the spatial coordinates linearly, which breaks translational invariance and gives rise to momentum relaxation.
Substituting the ansatz into~\eqref{eomphi}-\eqref{eomg}, we obtain the following equations:
\begin{equation}\label{eqphi}
\begin{split}
\frac{1}{\sqrt{BD}C}\left(\sqrt{\frac{D}{B}}C\phi'\right)'+\frac{\Omega}{C\sqrt{BD}}\frac{Z_2'(\phi)}{ Z_2(\phi)}\left((C^2+2 h^2 Z_2(\phi)^2)A_t'^2-h^2 BD\right)\\
-Z_1'(\phi)\left(\frac{Z_1(\phi)Z_2(\phi)^2}{\Omega C\sqrt{BD}}-1\right)-\frac{k^2 }{C}Y'(\phi)-V'(\phi)=0\,,
\end{split}
\end{equation}
\begin{equation}\label{eqmetric1}
\frac{D'C'}{DC}+\frac{1}{2}\frac{C'^2}{C^2}-{1\over 2}\phi'^2+B Z_1(\phi)\left(\frac{\Omega C\sqrt{BD}}{Z_1(\phi)Z_2(\phi)^2}-1\right)+\frac{\Omega B\sqrt{BD}h^2}{C}+\frac{k^2 B}{C}Y(\phi)+B V(\phi)=0\,,
\end{equation}
\begin{equation}\label{eqmetric2}
\frac{2C''}{C}-\left(\frac{B'}{B}+\frac{C'}{C}+\frac{D'}{D}\right)\frac{C'}{C}+\phi'^2=0\,,
\end{equation}
\begin{equation}\label{eqmetric3}
\frac{2D''}{D}-\frac{2C''}{C}-\left(\frac{B'}{B}-\frac{C'}{C}+\frac{D'}{D}\right)\frac{D'}{D}+\frac{B'C'}{BC}-2\Omega\sqrt{BD}\left(\frac{CA_t'^2}{D}+\frac{B h^2}{C}\right)-\frac{2 k^2 B}{C}Y(\phi)=0\,,
\end{equation}
\begin{equation}\label{eqA}
(\Omega (C^2+h^2 Z_2(\phi)^2) A_t')'=0\,,
\end{equation}
where for convenience we have introduced the function
\begin{equation}\label{omega}
\Omega(r)= \frac{Z_1(\phi) Z_2(\phi)^2}{\sqrt{(C^2+h^2 Z_2(\phi)^2)(BD-Z_2(\phi)^2 A_t'^2)}} \, .
\end{equation}
%

\section{DC Conductivities with a Finite Magnetic Field}
\label{ConductivitiesSection}
%

Next, we calculate the DC conductivities for our DBI model using the method developed in~\cite{Donos:2014uba,Blake:2014yla}\,\footnote{Other studies on the transport coefficients based on 
Einstein-Maxwell-like theories in the presence of a magnetic field can be found {\em e.g.} in~\cite{Blake:2015ina,Amoretti:2015gna,Blake:2015hxa,Kim:2015wba,Zhou:2015dha,Donos:2015bxe,Amoretti:2016cad,Ge:2016sel}.}.
We consider the following set of perturbations,
\begin{equation}\label{pertub}
\begin{split}
\delta g_{ti}=C(r) h_{ti}(r)\,,\quad \delta g_{ri}=C(r) h_{ri}(r)\,,\\
 \delta A_i=-E_i \,t+ a_i(r),\quad \delta\psi_1=\chi_1(x),\quad \delta\psi_2=\chi_2(x)\,,
\end{split}
\end{equation}
with $i=x,y$, and further simplify our analysis  by using diffeomorphisms to set
\begin {equation}\label{setU}
D(r)=\frac{1}{B(r)}\,.
\end{equation}
The vector equation~\eqref{eqA} can be immediately integrated, leading to the radially independent quantity
\begin{equation}
\label{charge}
\rho=\Omega (C^2+h^2 Z_2(\phi)^2) A_t'\,,
\end{equation}
which is nothing but the charge density $J^t$ in the dual field theory as defined in~\eqref{current}.
There are two constant fluxes that are provided by the perturbed vector equations,
\begin{equation}
\partial_r J^x(r)=\partial_r J^y(r)=0\,,
\end{equation}
where
\begin{eqnarray}
\label{JxJy1}
J^x(r)
&=&-\Omega CD (a_x'+h\,h_{ry}) -\frac{C^2 h_{tx}-h E_y Z_2^2}{C^2+h^2 Z_2^2} \, \rho \, ,\\
\label{JxJy2}
J^y(r) 
&=&-\Omega CD (a_y'-h\,h_{rx}) -\frac{C^2 h_{ty}+h E_x Z_2^2}{C^2+h^2 Z_2^2}\, \rho \, ,
\end{eqnarray}
are both currents in the dual field theory. Since they are conserved along the radial direction, they can be calculated anywhere in the bulk, 
with a particularly convenient choice being the horizon.

The perturbation equations coming from Einstein's equations~\eqref{eomg} are
\begin{equation}
h_{tx}''+\frac{2C'}{C}h_{tx}'+\Omega A_t'(a_x'+h\,h_{ry})-\frac{1}{CD}\left(k^2 Y+h^2\Omega\right)h_{tx}-\frac{h\Omega}{C D}E_y=0\,,
\end{equation}
\begin{equation}
h_{ty}''+\frac{2C'}{C}h_{ty}'+\Omega A_t'(a_y'-h\,h_{rx})-\frac{1}{CD}\left(k^2 Y+h^2\Omega\right)h_{ty}+\frac{h\Omega}{C D}E_x=0\,,
\end{equation}
\begin{equation}\label{htx}
k Y\chi_1'+\Omega h\, a_y'-\frac{\Omega C A_t'}{D}(E_x-h\,h_{ty})-\left(k^2 Y+\Omega h^2\right)h_{rx}=0\,,
\end{equation}
\begin{equation}\label{hty}
k Y\chi_2'-\Omega h\, a_x'-\frac{\Omega C A_t'}{D}(E_y+h\,h_{tx})-\left(k^2 Y+\Omega h^2\right)h_{ry}=0\,,
\end{equation}
while the axion equations~\eqref{eompsi} yield
\begin{equation}\label{chi1}
\chi_1''+\left(\frac{C'}{C}+\frac{D'}{D}+\frac{Y'\phi'}{Y}\right)(\chi_1'-k\, h_{rx})-k \,h_{rx}'=0\,,
\end{equation}
\begin{equation}\label{chi2}
\chi_2''+\left(\frac{C'}{C}+\frac{D'}{D}+\frac{Y'\phi'}{Y}\right)(\chi_2'-k\, h_{ry})-k \,h_{ry}'=0\,.
\end{equation}
Notice that~\eqref{chi1} and~\eqref{chi2} are implied by the other equations.

Since we are interested in a background geometry with a regular horizon at $r=r_h$, we have
\begin{equation}
\begin{split}
A_t&=A_t'(r_h) (r-r_h)+\dots,\\
D&=D'(r_h)(r-r_h)+\cdots=4\pi T(r-r_h)+\dots,
\end{split}
\end{equation}
while the constraint of regularity on the perturbation equations near $r_h$ demands the following expansions,
\begin{equation}\label{regular}
\begin{split}
a_i&=-\frac{E_i}{4\pi T}\log (r-r_h)+\dots,\quad h_{ti}=D h_{ri}+\dots,\\
\chi_1&=\chi_1(r_h)+\dots,\quad \chi_2=\chi_2(r_h)+\dots \;.
\end{split}
\end{equation}
The latter can be obtained by switching to the Eddington-Finklestein coordinate
\begin{equation}
v=t-\int^r \frac{1}{D(z)} dz=t-\frac{1}{4\pi T} \log(r-r_h)+\dots,
\end{equation}
where we have demanded that $v\rightarrow -\infty$ as $r\rightarrow r_h$.
Using the above regularity conditions, we extract the horizon data for $h_{tx}$ and $h_{ty}$ from~\eqref{htx} and~\eqref{hty},
\begin{equation}
h_{tx}(r_h)=-\frac{\rho\, E_x+h\, J^y}{k^2 C(r_h) Y(\phi(r_h))},\quad h_{ty}(r_h)=-\frac{\rho\, E_y-h\, J^x}{k^2 C(r_h) Y(\phi(r_h))} \, .
\end{equation}
Substituting the relations above into~\eqref{JxJy1} and~\eqref{JxJy2} and using~\eqref{regular}, we find
\begin{equation}\label{JJcurrent}
\begin{pmatrix}
   1+\frac{h^2\Omega}{k^2 Y}   &-\frac{C h\rho}{k^2 Y (C^2+h^2 Z_2^2)}    \\
  \frac{C h\rho}{k^2 Y (C^2+h^2 Z_2^2)}    &  1+\frac{h^2\Omega}{k^2 Y}   
\end{pmatrix}
\begin{pmatrix}
       J^x   \\
      J^y
\end{pmatrix}
=
\begin{pmatrix}
C\left(\Omega+\frac{\rho^2 }{k^2 Y (C^2+h^2 Z_2^2)}\right)      & h\rho\left(\frac{\Omega}{k^2 Y}+\frac{Z_2^2}{C^2+h^2 Z_2^2}\right)   \\
 -h\rho\left(\frac{\Omega}{k^2 Y}+\frac{Z_2^2}{C^2+h^2 Z_2^2}\right)     &  C\left(\Omega+\frac{\rho^2}{k^2 Y (C^2+h^2 Z_2^2)}\right)
\end{pmatrix}
\begin{pmatrix}
      E_x   \\
      E_y 
\end{pmatrix}\,,
\end{equation}
evaluated at the horizon $r=r_h$. 
In these expressions the function $\Omega$ introduced in~\eqref{omega} takes the form
\begin{equation}
\Omega=\frac{Z_2}{C^2+h^2 Z_2^2}\sqrt{\rho^2+Z_1^2 Z_2^2(C^2+h^2 Z_2^2)}\,.
\end{equation}
Finally, relating the two currents $J^x$ and $J^y$ in the matrix equation~\eqref{JJcurrent} to the electric fields $E_x$ and $E_y$ via
\begin{equation}
J^x = \sigma_{xx}\, E_x+\sigma_{xy}\, E_y\,,\quad
J^y= \sigma_{yx}\, E_x+\sigma_{yy}\, E_y \, ,
\end{equation}
the DC conductivities can be easily extracted and are given by
\begin{equation}\label{DC}
\begin{split}
\sigma_{xx}&=\sigma_{yy} = \frac{k^2 C Y \left[\Omega(h^2\Omega+k^2 Y)(C^2+h^2 Z_2^2)^2+C^2\rho^2\right]}{(h^2\Omega+k^2 Y)^2 (C^2+h^2 Z_2^2)^2+h^2 C^2 \rho^2}\, ,\\
\sigma_{xy}&=-\sigma_{yx} \\
&=\frac{h\rho[(h^2\Omega+k^2 Y)^2(h^2 Z_2^4+2 C^2Z_2^2)+(h^2\Omega+k^2 Y)C^4 \Omega+C^2\rho^2-C^2k^2Y(C^2 \Omega+k^2 Y Z_2^2)]}{(h^2\Omega+k^2 Y)^2 (C^2+h^2 Z_2^2)^2+h^2 C^2 \rho^2} \,.
\end{split}
\end{equation}
The conductivity matrix is controlled by four functions, the three scalar couplings $Z_1, Z_2, Y$ and the component $C$ of the bulk metric. 
All four are functions of the holographic radial coordinate $r$ and in  (\ref{DC}) are evaluated at the horizon $r=r_h$. Moreover, since $r_h$ is in general a function of temperature $T$, 
the matrix (\ref{DC}) is implicitly temperature-dependent, while the dependence on the remaining 
scales in the system -- the magnetic field $h$, the 
strength of momentum dissipation $k$ and the charge density $\rho$ -- is explicitly visible.
We should note that our results for $\sigma_{xx}$ overlap with those obtained recently in \cite{Blake:2017qgd}.

From the expressions (\ref{DC}) we can then extract the inverse Hall angle,
\begin{equation}
\cot \Theta_H=\frac{\sigma_{xx}}{\sigma_{xy}},
\end{equation}
and the resistivity matrix by inverting the conductivity matrix,
\begin{equation}\label{resis}
R_{xx}=R_{yy}=\frac{\sigma_{xx}}{\sigma_{xx}^2+\sigma_{xy}^2}\,,\quad R_{xy}=-R_{yx}=-\frac{\sigma_{xy}}{\sigma_{xx}^2+\sigma_{xy}^2}\, .
\end{equation}
From now on all functions will be understood to be evaluated at the horizon, but for convenience we will omit the explicit dependence on $r_h$. 
Since the general formulae for $\sigma_{ij}$ and $R_{ij}$ are quite cumbersome, we consider first some simple limiting cases.

\subsection{Weak momentum dissipation}
A simple case to consider is that of slow momentum relaxation, \emph{i.e.} small $k$.
As a consistency check, we first look at the limit $k\rightarrow 0$, which corresponds to no momentum dissipation at all. The conductivity tensor then reduces to
\begin{equation}\label{weakk0}
\sigma_{xx}=\sigma_{yy}=0,\quad \sigma_{xy}=-\sigma_{yx}=\frac{\rho}{h} \, ,
\end{equation}
and is independent of the temperature as well as the details of the theory we are working with. 
This can be understood as a generic consequence of Lorentz invariance when $k\rightarrow 0$, 
and agrees with the Hall conductivity result of~\cite{Hartnoll:2007ai}.
Including the leading and subleading corrections coming from momentum dissipation, we find 
%
\begin{equation}
\begin{split}
\sigma_{xx}&=\sigma_{yy}=\frac{ C}{h^2}k^2 Y-\frac{C\Omega(C^2+h^2 Z_2^2)^2}{h^4\Omega^2(C^2+h^2 Z_2^2)^2+h^2 C^2 \rho^2}(k^2 Y)^2+ \ldots \,,\\
\sigma_{xy}&=-\sigma_{yx}=\frac{\rho}{h}-\frac{\rho}{h}\frac{C^2 (C^2+h^2 Z_2^2)}{h^4\Omega^2(C^2+h^2 Z_2^2)^2+h^2 C^2 \rho^2}(k^2 Y)^2+ \ldots \,.
\end{split}
\end{equation}
As expected, the matrix components are now sensitive to the detailed structure of the model, and are temperature dependent through the implicit dependence on $r_h$.

\subsection{Vanishing magnetic field}
In the absence of magnetic field, $\sigma_{xy}=0$, and the DC conductivity reduces to the simple expression
\begin{equation}\label{dcnoh}
\sigma_{DC}=\sigma_{xx}=Z_2\sqrt{Z_1^2 Z_2^2+\frac{ \rho^2}{C^2}}+\frac{\rho^2}{k^2 Y C}=Z_2\sqrt{Z_1^2 Z_2^2+\frac{16\pi^2 \rho^2}{s^2}}+\frac{4\pi \rho^2}{k^2 Y s}\,,
\end{equation}
where $s=4\pi \,C(r_h)$ is the entropy density\,\footnote{In the action~\eqref{action} we have used units with $\frac{1}{16\pi G_N}=1$, where $G_N$ is Newton's constant. So the entropy density by definition is $s=\frac{C(r_h)}{4 G_N}=4\pi C(r_h)$.}. 
As seen in a number of cases in the literature, 
the DC conductivity can be interpreted  \cite{Davison:2015bea} as being composed of two physically distinct and independent pieces: 
a coherent contribution $\sigma_{DC}^{diss}$ due to momentum relaxation for the charge carriers in the system, 
and an incoherent contribution, known as the charge conjugation symmetric term $\sigma_{DC}^{ccs}$, which is independent of the charge density $\rho$.
In the absence of magnetic field, there are examples showing that the DC conductivity consists of such two distinct terms, simply added together.
However, more generally the contributions can combine to form the DC conductivity in a rather non-trivial fashion.
Indeed, notice that in~\eqref{dcnoh} we do not have a clean separation between $\sigma_{DC}^{ccs}$ and terms dissipating momentum for charge carriers. 
The first contribution in the square root persists at zero charge density,  \emph{i.e.}  the charge conjugation symmetric term is given by 
$\sigma_{DC}^{ccs}=Z_1 Z_2^2$. 
The other two terms are associated with the charge density $\rho$  and are due to momentum dissipation effects.
Thus, here we have given an explicit realization of a setup in which there is no simple separation between 
$\sigma_{DC}^{ccs}$ and $\sigma_{DC}^{diss}$.

\subsection{Vanishing charge density}
The DC resistivity in the absence of charge density reads
\begin{equation}\label{dcnorho}
R_{DC}=R_{xx}=\frac{1}{Z_1 Z_2^2}\sqrt{1+\frac{Z_2^2}{C^2}h^2}+\frac{h^2}{k^2 Y C}=\frac{1}{Z_1 Z_2^2}\sqrt{1+\frac{16\pi^2 Z_2^2}{s^2}h^2}+\frac{4\pi h^2}{k^2 Y s}\,,
\end{equation}
which falls into  the charge conjugation regime, since the charge density is vanishing. It should be pointed out that charge fluctuations still exist at zero charge density, and  it would seem the incoherent conductivity should be identified as being due to diffusion of charge fluctuations\,\footnote{We would like to thank the referee for clarifying these points.}.
Notice the similarity of the structure of this result with that of (\ref{dcnoh}). In particular, we have $\sigma_{xy} = R_{xy}=0$ because $\rho=0$. 
In contrast, in the case with vanishing magnetic field the theory is parity symmetric, which requires the Hall conductivity to vanish for any value of charge density$\rho$.

\subsection{Strong momentum dissipation limit}
Next, we consider the case in which the momentum dissipation $\sim k$ is dominant compared to the other scales in the system. 
Working to leading order in the strong momentum dissipation limit, 
we obtain the conductivities
\begin{equation}
\begin{split}
&\sigma_{xx}=\sigma_{yy}=\Omega C-\frac{C(\Omega^2 h^2(C^2+h^2 Z_2^2)^2-C^2\rho^2)}{(C^2+h^2 Z_2^2)^2}\frac{1}{k^2 Y}+ \ldots \,,\\
& \sigma_{xy}=-\sigma_{yx}=\frac{h \rho Z_2^2}{C^2+h^2 Z_2^2}+\frac{2 C^2 h\rho \Omega}{C^2+h^2 Z_2^2}\frac{1}{k^2 Y}+ \ldots \,,
\end{split}
\end{equation}
and the corresponding resistivities
\begin{equation}
\begin{split}
R_{xx}=&R_{yy}=\frac{C}{Z_2}\frac{\sqrt{\rho^2+Z_1^2 Z_2^2(C^2+h^2 Z_2^2)}}{\rho^2+C^2 Z_1^2 Z_2^2}-\\
&\frac{C[\rho^2(\rho^2+C^2 Z_1^2 Z_2^2)+h^2 Z_1^2 Z_2^4(\rho^2-C^2 Z_1^2 Z_2^2)]}{Z_2^2(\rho^2+C^2 Z_1^2 Z_2^2)^2}\frac{1}{k^2 Y}+ \ldots \,,\\
R_{xy}=&-R_{yx}=-\frac{h \rho}{\rho^2+C^2 Z_1^2  Z_2^2}-\\
&\frac{2h\rho C^2 Z_1^2 Z_2\sqrt{\rho^2+Z_1^2 Z_2^2(C^2+h^2 Z_2^2)}}{Z_2^2(\rho^2+C^2 Z_1^2 Z_2^2)^2}\frac{1}{k^2 Y}+  \ldots \,.
\end{split}
\end{equation}
We focus on the conductivities at leading order, which are given by
\begin{equation}
\label{strongkresult}
\begin{split}
&\sigma_{xx}=\sigma_{yy}=\Omega C=\frac{Z_2 C}{C^2+h^2 Z_2^2}\sqrt{\rho^2+Z_1^2 Z_2^2(C^2+h^2 Z_2^2)}\,,\\
& \sigma_{xy}=-\sigma_{yx}=\frac{h \rho Z_2^2}{C^2+h^2 Z_2^2} \,.
\end{split}
\end{equation}
The inverse Hall angle reads
\begin{equation}
\label{strongkhall}
\cot \Theta_H=\frac{\sigma_{xx}}{\sigma_{xy}}=\frac{C}{h\rho Z_2}\sqrt{\rho^2+Z_1^2 Z_2^2(C^2+h^2 Z_2^2)}\,,
\end{equation}
and the in-plane resistivity
\begin{equation}\label{rhoxx}
R_{DC}=R_{xx}=\frac{C}{Z_2}\frac{\sqrt{\rho^2+Z_1^2 Z_2^2(C^2+h^2 Z_2^2)}}{\rho^2+C^2 Z_1^2 Z_2^2}\,.
\end{equation}
Interestingly, we find that these expressions are precisely the same as the ones which were obtained in the probe DBI case~\cite{Kiritsis:2016cpm}, using a different approach. 

This can be understood as follows. When the momentum dissipation is strong enough, the contribution to the geometry coming
from the DBI sector is negligible compared to that of the axionic sector. 
Thus, in this case the background geometry is seeded by the axions, and the dynamics of the U(1) gauge field 
can be captured by treating it as a probe around the resulting geometry.
This can be easily seen from the background equations~\eqref{eqphi}-\eqref{eqmetric3}. 
When the terms coming from the DBI action are negligible compared to the axionic terms, we obtain a closed system which only involves the axions as well as $\phi$ coupled to gravity,
\begin{equation}\label{phiaxion}
\begin{split}
\frac{1}{\sqrt{BD}C}\left(\sqrt{\frac{D}{B}}C\phi'\right)'-\frac{k^2 }{C}Y'(\phi)-V'(\phi)=0\,,
\end{split}
\end{equation}
\begin{equation}\label{metric1axion}
\frac{D'C'}{DC}+\frac{1}{2}\frac{C'^2}{C^2}-{1\over 2}\phi'^2+\frac{k^2 B}{C}Y(\phi)+B V(\phi)=0\,,
\end{equation}
\begin{equation}\label{metric2axion}
\frac{2C''}{C}-\left(\frac{B'}{B}+\frac{C'}{C}+\frac{D'}{D}\right)\frac{C'}{C}+\phi'^2=0 \, ,
\end{equation}
\begin{equation}\label{metric3axion}
\frac{2D''}{D}-\frac{2C''}{C}-\left(\frac{B'}{B}-\frac{C'}{C}+\frac{D'}{D}\right)\frac{D'}{D}+\frac{B'C'}{BC}-\frac{2 k^2 B}{C}Y(\phi)=0\,.
\end{equation}
The gauge field $A_t$ can then be determined from~\eqref{eqA}. 

As was shown in~\cite{Kiritsis:2016cpm,Gouteraux:2014hca}, the coupled equations of motion~\eqref{phiaxion}-\eqref{metric3axion} admit IR hyperscaling scaling violating geometries, 
\begin{eqnarray}\label{axionfinitT}
ds^2&=&r^\theta\left(-f(r)\frac{dt^2}{r^{2z}}+\frac{L^2 dr^2}{r^2 f(r)}+\frac{dx^2+dy^2}{r^2}\right)\,,\\
\phi&=&\kappa\,\ln(r),\quad \psi^1=k\,x,\quad \psi_2=k\,y\,,\nonumber
\end{eqnarray}
with
\begin{eqnarray}\label{solutionhsv}
&& f(r)=1-\left(\frac{r}{r_h}\right)^{2+z-\theta}\sp z=\frac{\alpha^2-\eta^2+1}{\alpha(\alpha+\eta)}\sp \theta=\frac{2\eta}{\alpha},\quad \kappa=-\frac{2}{\alpha}\,,\nonumber\\
&&L^2=\frac{(z+2-\theta)(\theta-2z)}{V_0 }\sp k^2 L^2=2(z-1)(z+2-\theta)\,,\\
&& \eta=\pm\frac{\theta}{\sqrt{(\theta-2)(\theta+2-2z)}}\sp \alpha=\pm\frac{2}{\sqrt{(\theta-2)(\theta+2-2z)}}\,,\nonumber
\end{eqnarray}
when the dilaton couplings $V$ and $Y$ are approximated by exponentials in the IR,
\begin{equation}\label{asympvy }
V(\phi)\sim -V_0 \,e^{\eta\, \phi},\quad Y(\phi)\sim e^{\alpha\,\phi}\,,
\end{equation}
with $\eta, \alpha$ constants. 
In order to have a well defined geometry and a resolvable singularity one should take into account the Gubser's physicality criterion~\cite{Gubser:2000nd,Charmousis:2010zz}, which restricts the range of the scaling exponents $\{z, \theta\}$ appearing in~\eqref{solutionhsv}. 
In particular, the allowed parameter range is given by
\begin{equation}
\begin{split}
& \text{IR} \quad r\rightarrow 0: \; \; \; [z\leqslant 0, \theta>2],\quad [0<z<1, \theta>z+2]\,,\\
&\text{IR} \quad r\rightarrow \infty : \; [1<z\leqslant 2, \theta<2z-2],\quad [z>2, \theta<2]\, ,
\end{split}
\end{equation}
depending on the location of the IR.
It was also shown in~\cite{Kiritsis:2016cpm} that by setting
\begin{equation}
\frac{C(r_h)}{Z_2(\phi(r_h))}\sim T,\quad Z_1(\phi(r_h))Z_2(\phi(r_h))^2\sim\frac{1}{T}\,,
\end{equation}
where $T$ is the temperature, one can obtain the scaling behavior
\begin{equation}
R_{DC}\sim \sqrt{a T^2+h^2}\,,
\end{equation}
with $a$ a constant which depends on the details of the action. 
The main point to note for this case is that for appropriate choices of parameters it is possible to reproduce 
the in-plane resistance~\eqref{experiment}.  The anomalous temperature dependence of the resistivity and Hall angle of the cuprate strange metals has recently been realized in this setup~\cite{Blauvelt:2017koq}. The backreacted DBI case, however, leads to a much richer transport behavior, as we will see next.

\section{Magnetic-Field-Induced Metal-Insulator Transition}
\label{metalinsulator}

If we choose the dilaton field $\phi$ to be trivial, the background black brane geometry can be solved exactly.
Even in this simple case the physics is still quite rich, and we find a finite-temperature transition -- or crossover -- from metallic to insulating behavior, induced by the magnetic field.

We take the couplings to be of the form\,\footnote{A class of exact solutions for the DBI theory without axions have been studied in~\cite{Kundu:2016oxg}.}
\begin{equation}\label{adstheory}
Z_1=z_1\,,\quad Z_2=Y=1\,,\quad V=-V_0,\quad \phi=0\,,
\end{equation}
where $z_1$ and $V_0$ are positive constants. Once again we set $D(r)=1/B(r)$. The metric function $C(r)$ is then found by solving~\eqref{eqmetric2}, and is given by
\begin{equation}
C(r)=r^2.
\end{equation}
Here we have chosen the AdS boundary to be at $r\rightarrow \infty$.
The remaining (non-trivial) equations of motion are then
\begin{eqnarray}
&& A_t'-\frac{\rho}{z_1\sqrt{r^4+\frac{\rho^2+h^2 z_1^2}{z_1^2}}}=0 \, , \\
&& r D'+D-\frac{r^2}{2}(V_0+z_1)+\frac{1}{2}k^2+\frac{z_1}{2}\sqrt{r^4+\frac{\rho^2+h^2 z_1^2}{z_1^2}}=0 \, ,\label{eomsads} \\
&& D''-\frac{2}{r^2}D-\frac{1}{r^2}k^2-\frac{\rho^2+h^2 z_1^2}{r^2 z_1\sqrt{r^4+\frac{\rho^2+h^2 z_1^2}{z_1^2}}}=0 \, .
\end{eqnarray}
We find that the last equation is implied by the second one.
Solving~\eqref{eomsads}, we obtain
\begin{equation}
\begin{split}
D(r)=\frac{r^2}{6}(V_0+z_1)-&\frac{z_1}{6}\sqrt{r^4+\frac{h^2 z_1^2+\rho ^2}{z_1^2}}-\frac{1}{2}k^2-\frac{M}{r}\\
-&\frac{1}{3}\sqrt{h^2 z_1^2+\rho ^2} \;
   _2F_1\left(\frac{1}{4},\frac{1}{2};\frac{5}{4};-\frac{r^4 z_1^2}{\rho ^2+h^2 z_1^2}\right) ,
 \end{split}
\end{equation}
where $M$ corresponds to the mass of the black brane and is determined by the location of the horizon $r_h$ via $D(r_h)=0$. 
The U(1) gauge field is given by
\bea
A_t(r) &=& \int_{r_h}^r \frac{\rho}{\alpha\sqrt{u^4+\frac{\rho^2+h^2 z_1^2}{z_1^2}}} \, du \nonumber \\
&=& c_1+\rho\, r \sqrt{\frac{1}{h^2 z_1^2+\rho ^2}} \;
   _2F_1\left(\frac{1}{4},\frac{1}{2};\frac{5}{4};-\frac{r^4 z_1^2}{\rho ^2+h^2 z_1^2}\right) ,
\eea
with the constant $c_1$ given by requiring as usual that the gauge field vanishes at the horizon, $A_t(r_h)=0$.
Finally, the temperature associated with the black brane geometry takes the form
\begin{equation}
\label{adsT}
T=\frac{D'(r_h)}{4\pi}=\frac{V_0+z_1}{8\pi}r_h-\frac{1}{8\pi r_h}k^2-\frac{z_1}{8\pi r_h}\sqrt{r_h^4+\frac{\rho^2+h^2 z_1^2}{z_1^2}}\, ,
\end{equation}
and the entropy density reads
\begin{equation}
s=4\pi r_h^2\,.
\end{equation}
By making use of~\eqref{adsT} to express the location of the horizon in terms of $T$, one can find 
the temperature dependence of the conductivity matrix~\eqref{DC} as well as the resistivity matrix~\eqref{resis}, which 
of course also depends on the magnetic field $h$, the charge density $\rho$ and 
the momentum dissipation parameter $k$.

It is interesting to ask whether the black brane solution we just presented leads  to metallic or insulating behavior. 
To this end, we are going to adopt the following working definition of a metal versus an insulator,
\begin{equation}\label{criterion}
\text{Metal}:\quad \frac{d R_{xx}}{d T}>0\,,\quad\quad \text{Insulator}:\quad \frac{d R_{xx}}{d T}<0 \, ,
\end{equation}
and inspect the temperature dependence of the conductivities.
We will focus on cases with finite momentum dissipation, since in the limit
$k\rightarrow 0$ shown in~\eqref{weakk0} the conductivity is quite simple, due to Lorentz invariance. 
For simplicity and without loss of generality, from now on we fix our theory parameters to be
\begin{equation}
z_1=1,\quad V_0=6 \, .
\end{equation}
We start by considering two simple cases which correspond to, respectively, vanishing magnetic field and charge density.
The former turns out to be associated with metallic behavior, while the latter with insulating.
We then look at the more generic situation, in which both $h$ and $\rho$ are non-zero, and find a finite temperature crossover between the two types of behavior.

\subsection{Vanishing magnetic field}
We examine first the case in which the magnetic field is absent.
The Hall part of the conductivity is zero, and the resistivity can be obtained from~\eqref{dcnoh},
\begin{equation}
R_{DC}=R_{xx}=1/\sigma_{xx}=\frac{k^2 r_h^2}{\rho^2+k^2\sqrt{r_h^4+\rho^2}}\,,
\end{equation}
\begin{figure}[ht!]
\begin{center}
\includegraphics[width=.45\textwidth]{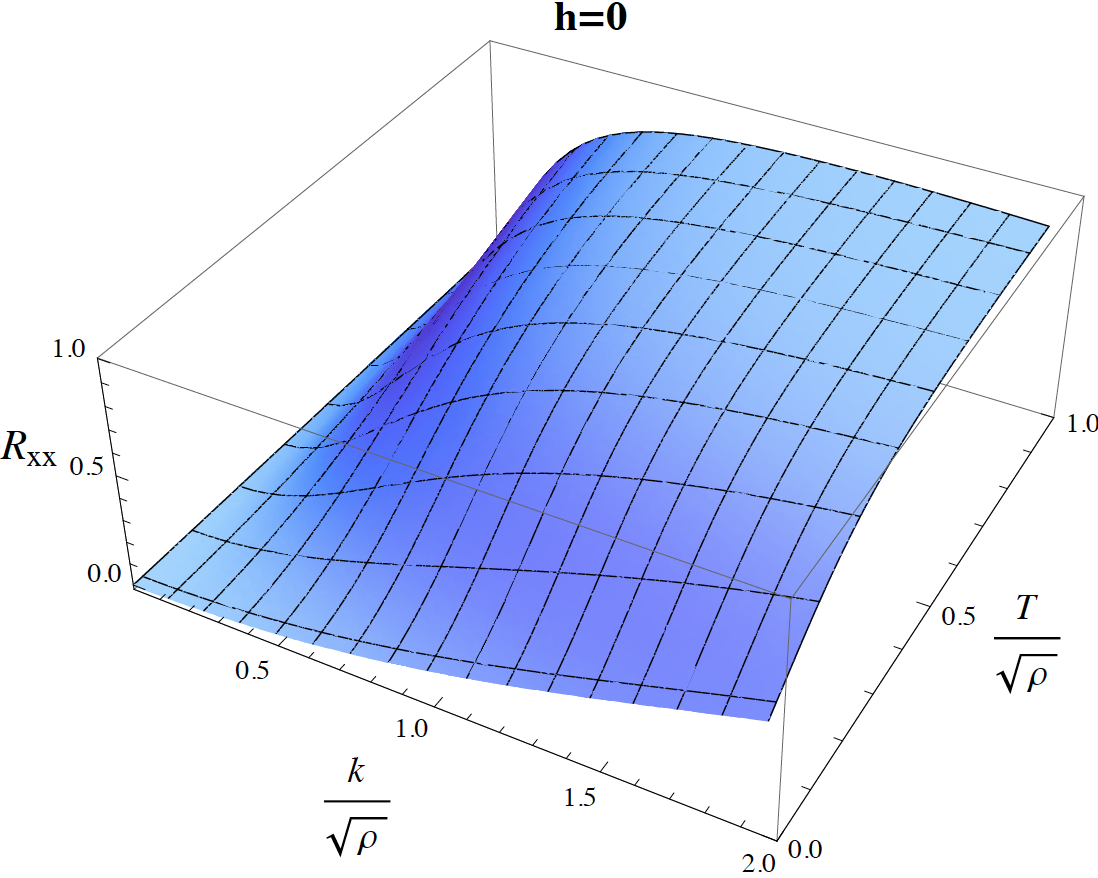}\quad
\includegraphics[width=.45\textwidth]{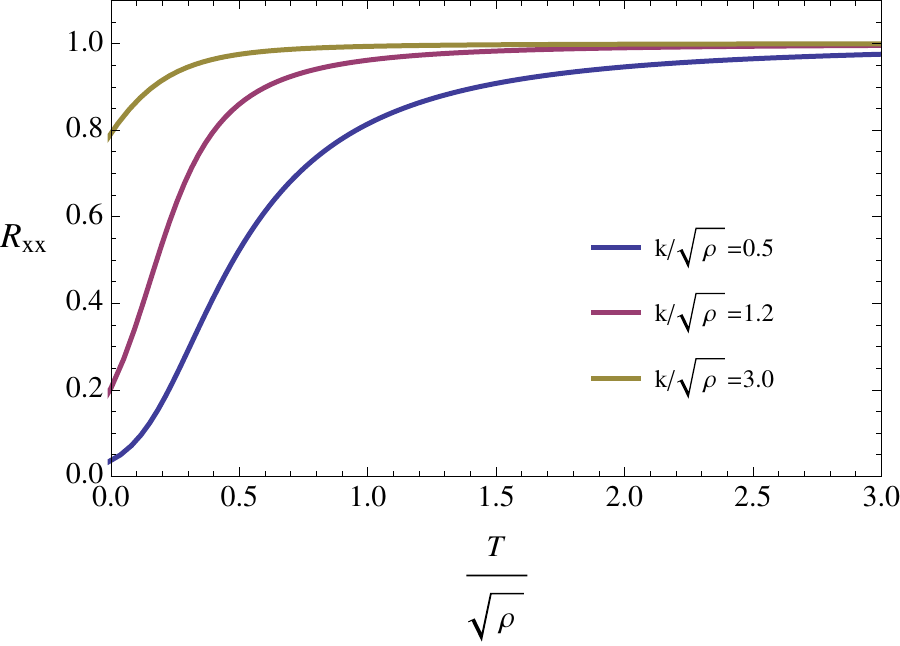}
\caption{The DC resistivity $R_{xx}$ when $h=0$ as a function of $T/\sqrt{\rho}$ and $k/\sqrt{\rho}$. 
Moving from top to bottom, the curves in the right panel correspond to decreasing values of $k/\sqrt{\rho}$.}
\label{fig:mrdcho}
\end{center}
\end{figure}
For a fixed value of $k/\sqrt{\rho}$,\,\footnote{Without loss of generality, we will assume $\rho\geqslant 0$ from now on.} the resistivity $R_{DC}$ increases monotonically with increasing temperature, as shown in Figure~\ref{fig:mrdcho}.
Thus, according to the criterion~\eqref{criterion}, the resulting behavior is metallic. 
The curves displayed in the right panel of Figure~\ref{fig:mrdcho} correspond to, from top to bottom, decreasing values of $k/\sqrt{\rho}$.
We therefore see that by lowering the amount $k$ of momentum dissipation one also decreases the resistivity, with the effect being especially pronounced at low $T$. What this indicates is that as $k \rightarrow 0$ we should recover a divergent conductivity, which is expected from the fact that we would be approaching the regime of no momentum dissipation.

\subsection{Vanishing charge density}
In the case with vanishing charge density, the component $\sigma_{xy}$ is also zero. 
The resistivity can now be obtained from~\eqref{dcnorho},
\begin{equation}
R_{DC}=R_{xx}=\frac{h^2+k^2\sqrt{r_h^4+h^2}}{k^2 r_h^2} \, .
\end{equation}
\begin{figure}[ht!]
\begin{center}
\includegraphics[width=.45\textwidth]{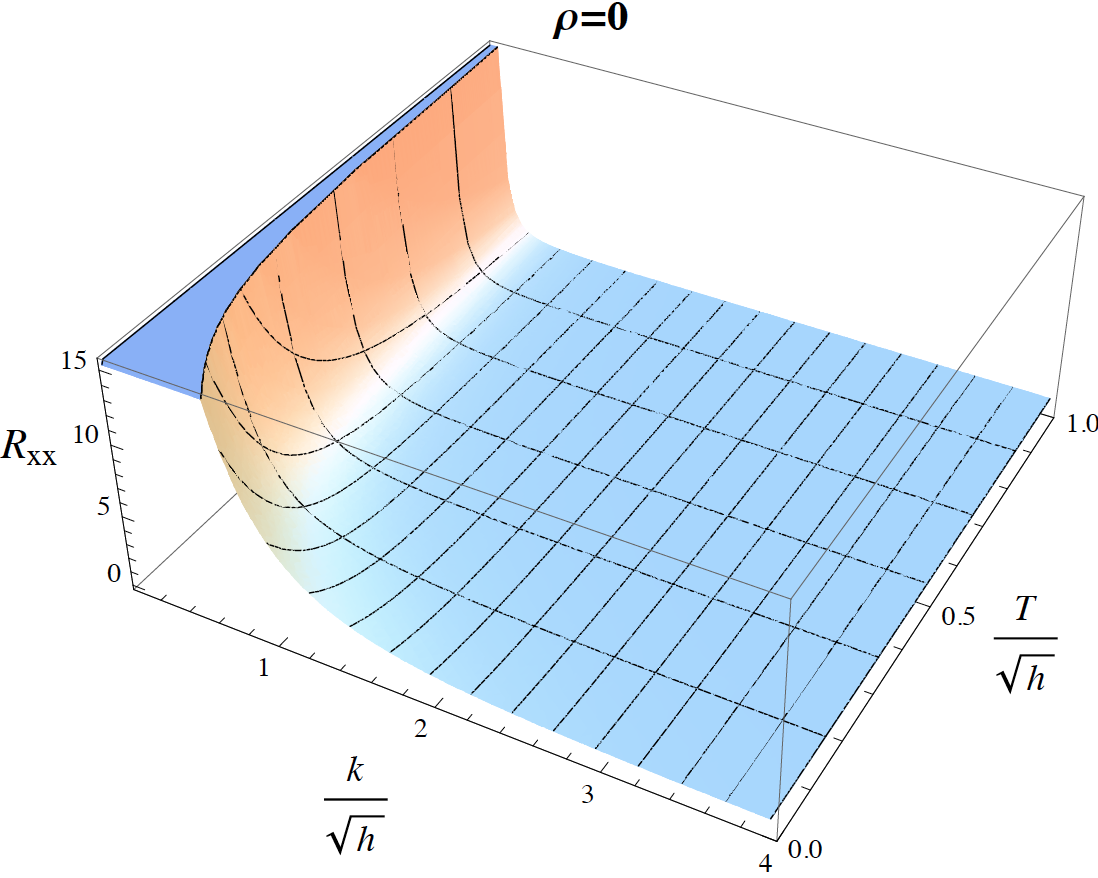}\quad
\includegraphics[width=.45\textwidth]{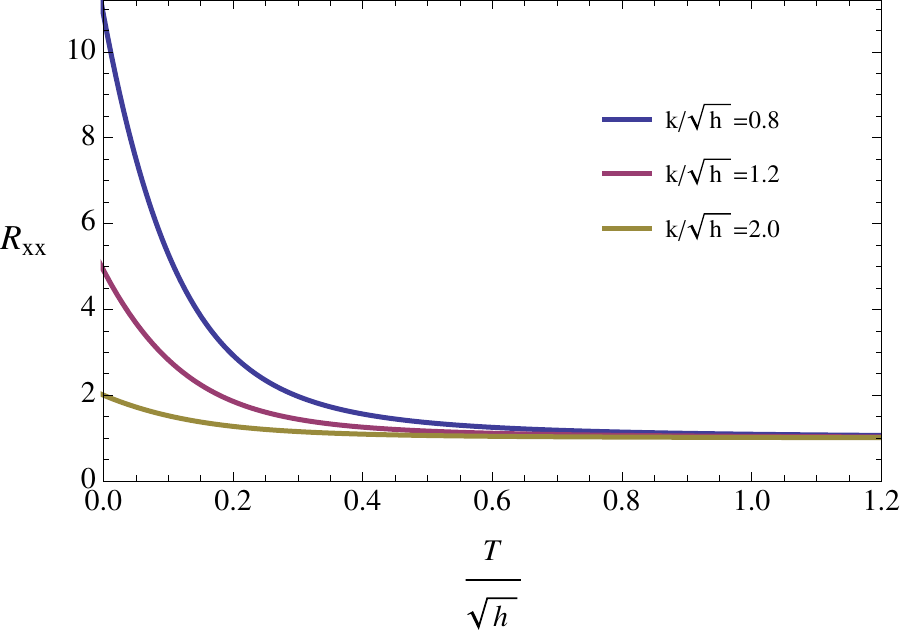}
\caption{The DC resistivity $R_{xx}$ at vanishing charge density as a function of $T/\sqrt{h}$ and $k/\sqrt{h}$. 
In the right panel, from top to bottom the ratio $k/\sqrt{h}$ increases.}
\label{fig:mrdcrho0}
\end{center}
\end{figure}
For a fixed value of $k/\sqrt{h}$, the quantity $R_{DC}$ decreases monotonically as the temperature increases, as can be seen from Figure~\ref{fig:mrdcrho0}. 
According to~\eqref{criterion}, this corresponds to an insulating like behavior. 
Moreover, the curves in the right panel of Figure~\ref{fig:mrdcrho0} show that the smaller the ratio $k/\sqrt{h}$, the larger the resistivity, with the enhancement more pronounced at low temperatures.
We wonder whether this effect is entirely model dependent, or whether it could be a feature of the role of disorder or momentum dissipation on insulating phases.

\subsection{Magnetotransport at finite magnetic field and charge density}
\begin{figure}[ht!]
\begin{center}
\includegraphics[width=.45\textwidth]{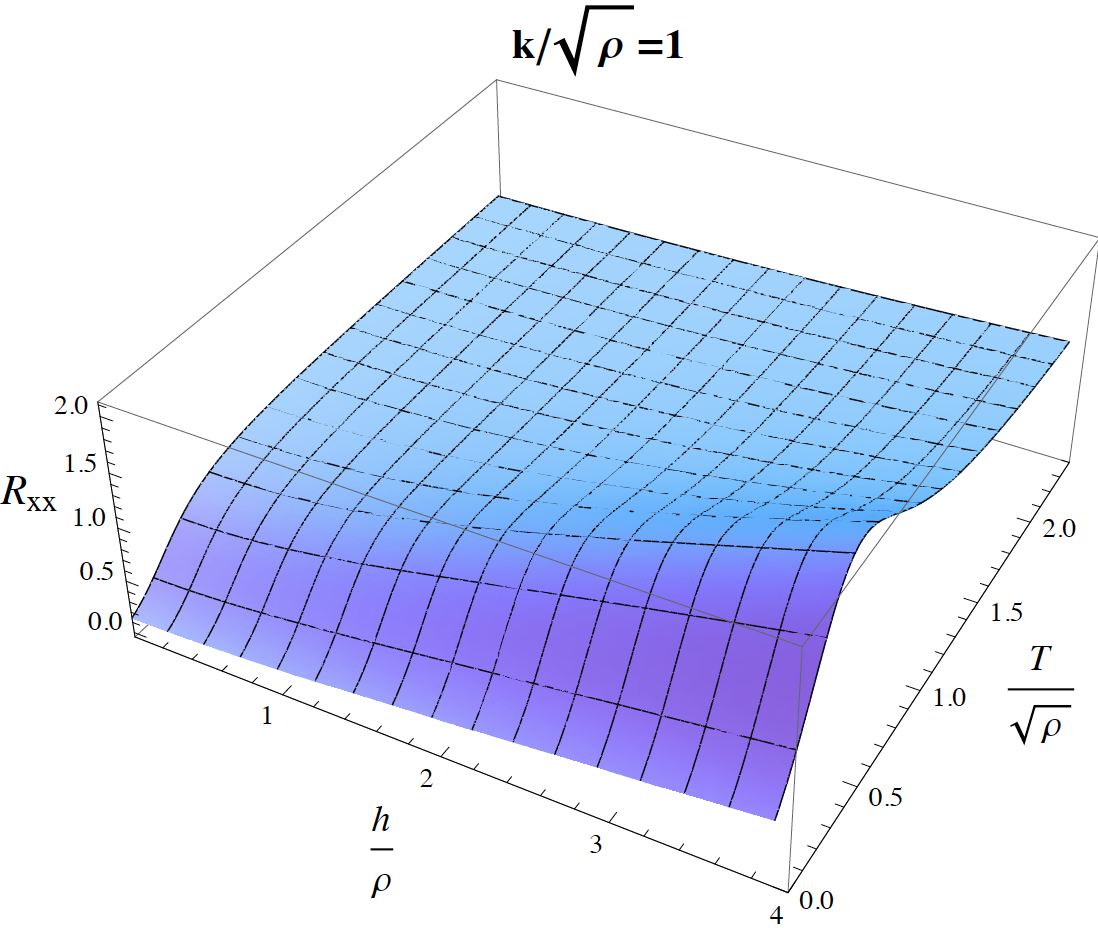}\quad
\includegraphics[width=.45\textwidth]{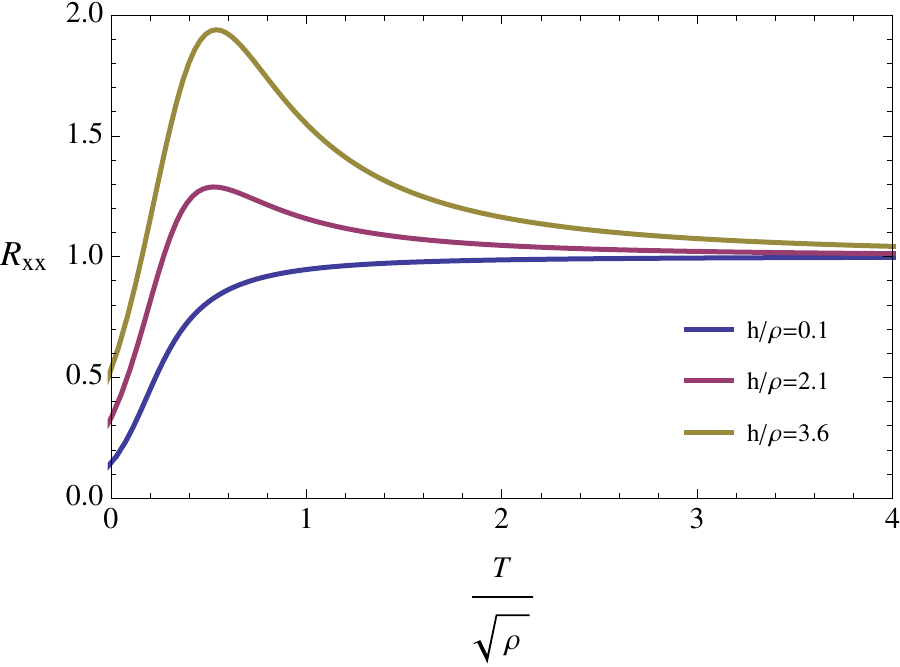}
\caption{The resistance $R_{xx}$ at finite charge density as a function of $h/\rho$ and $T/\sqrt{\rho}$. 
We choose the momentum dissipation parameter $k/\sqrt{\rho}=1$. In the right panel, the curves from top to bottom correspond to decreasing values of $h/\rho$.}
\label{fig:mrdc}
\end{center}
\end{figure}

We move on to the more generic case in which both $h$ and $\rho$ are non-zero, which is significantly more 
complex. There is now a non-trivial Hall component to the conductivities, and the general resistivity components are
given by
\begin{eqnarray}
R_{xx}&=&k^2 \, r_h^2 \, \frac{\rho ^2 Q^2 \left(Q^2-k^4\right)+k^2\sqrt{Q ^2+r_h^4}
   \left(\rho ^2 \left(k^4-Q ^2\right)+k^4 r_h^4\right)+k^4 r_h^4(h^2-\rho^2 )}{\rho ^4 \left(Q^2-k^4\right)^2+2 k^4 \rho ^2 r_h^4 \left(h^2+k^4-\rho^2\right)+k^8 r_h^8},  \nn \\
R_{yx}&=&h \, \rho \, \frac{2k^6  r_h^4 \sqrt{Q^2+r_h^4}+\rho^2 \left(Q^2-k^4\right)^2+k^4 r_h^4 \left(h^2+k^4-3 \rho ^2\right)}{\rho ^4 \left(Q^2-k^4\right)^2+2 k^4 \rho ^2 r_h^4 \left(h^2+k^4-\rho^2\right)+k^8 r_h^8},
\end{eqnarray}
where we have introduced $Q^2=\rho^2+h^2$. The inverse Hall angle reads
\begin{equation}
\begin{split}
\cot \Theta_H&=\frac{\sigma_{xx}}{\sigma_{xy}}=\frac{R_{xx}}{R_{yx}}\\
&=\frac{k^2 r_h^2}{h\rho}\frac{\rho ^2 Q^2 \left(Q^2-k^4\right)+k^2\sqrt{Q ^2+r_h^4}
   \left(\rho ^2 \left(k^4-Q ^2\right)+k^4 r_h^4\right)+k^4 r_h^4(h^2-\rho^2 )}{2k^6  r_h^4 \sqrt{Q^2+r_h^4}+\rho^2 \left(Q^2-k^4\right)^2+k^4 r_h^4 \left(h^2+k^4-3 \rho ^2\right)}.
   \end{split}
\end{equation}
We display the behavior of the in-plane resistance $R_{xx}$ in Figure~\ref{fig:mrdc} at the momentum dissipation parameter $k/\sqrt{\rho}=1$.  We find the following features:
\begin{itemize}
  \item $h<\rho$: $R_{xx}$ increases monotonically as one increases the temperature, corresponding to metallic behavior.
  \item $h>\rho$: As the temperature increases, $R_{xx}$ first rises, then reaches a maximum at a certain ratio $T_0/\sqrt{\rho}$, and then decreases monotonically. 
The value of $T_0/\sqrt{\rho}$ depends on $k/\sqrt{\rho}$ and $h/\rho$. We have metallic behavior at low temperatures and insulating at high temperatures. 
Thus, this can be thought of as a metal-insulator transition -- or crossover -- induced by the magnetic field. 
\end{itemize}

\begin{figure}[ht!]
\begin{center}
\includegraphics[width=.45\textwidth]{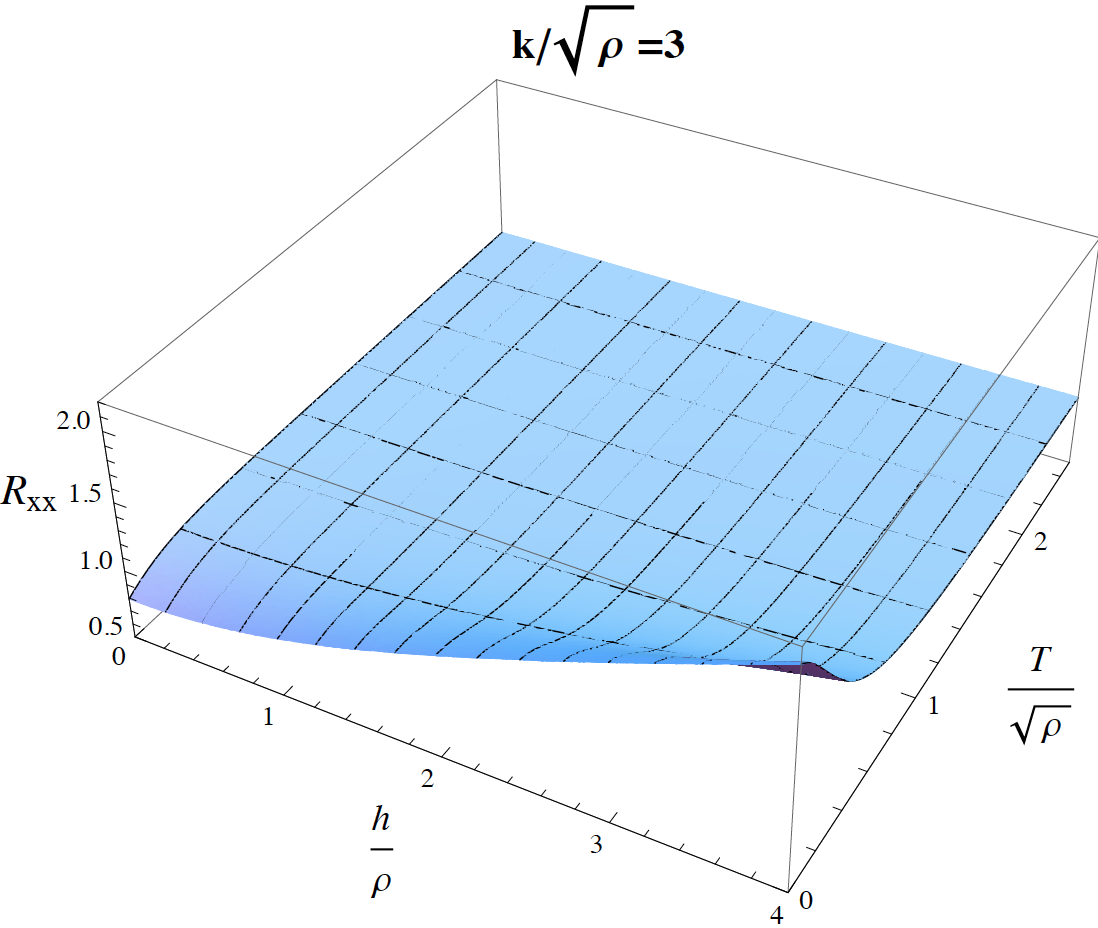}\quad
\includegraphics[width=.45\textwidth]{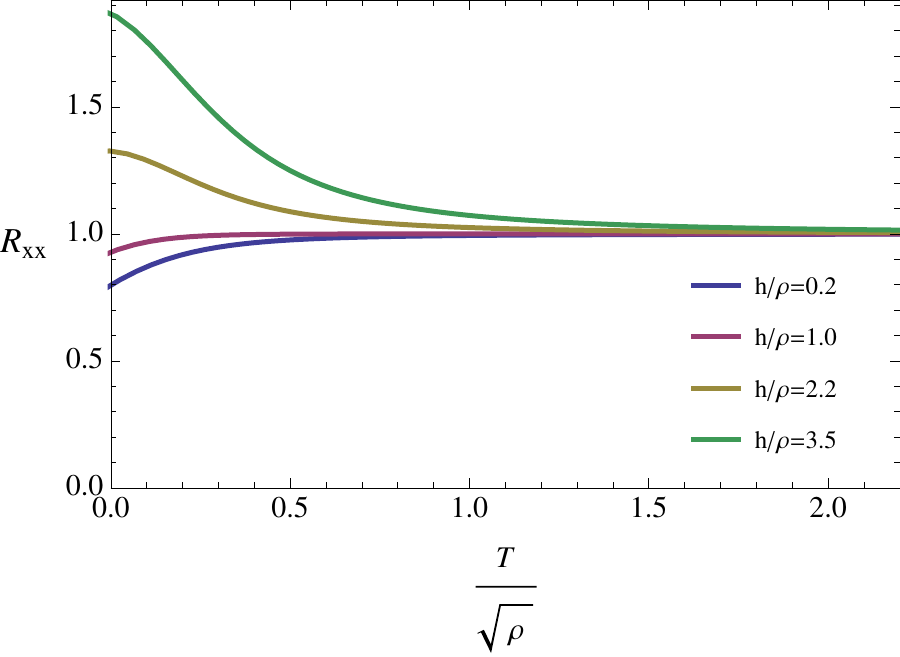}
\caption{The resistance $R_{xx}$ at finite charge density versus $h/\rho$ and $T/\sqrt{\rho}$. 
We choose the momentum dissipation parameter $k/\sqrt{\rho}=3$. In the right panel, the curves from top to bottom correspond to decreasing values of $h/\rho$.}
\label{fig:mrdcl}
\end{center}
\end{figure}

We also display the resistance $R_{xx}$ for a larger value of the momentum dissipation parameter, $k/\sqrt{\rho}=3$, in Figure~\ref{fig:mrdcl}. 
The temperature dependence of $R_{xx}$ is similar to that in the previous case when $h<\rho$. 
However, the non-monotonic behavior at large values of the magnetic field disappears and $R_{xx}$ decreases monotonically as one increases the temperature, which is reminiscent of an insulating behavior. Note that the change in the behavior of the resistivity is once again induced by the magnetic field. 
Metal-insulator transitions or crossovers have been studied using other gravity setups, see, \emph{e.g.}~\cite{Donos:2012js,Ling:2015epa,Cai:2015wfa,Baggioli:2016oqk,Ling:2016dck,Cremonini:2016avj}.

The threshold value for the magnetic field, $h/\rho=1$, can be understood in the following way. 
Consider the high temperature limit $T\gg (k,\sqrt{\rho},\sqrt{h})$ in which $T$ is the dominant scale in the problem. 
In this limit at leading order the temperature~\eqref{adsT} is given by the simple expression
\begin{equation}
T= \frac{V_0}{8\pi}r_h = \frac{3}{4\pi}r_h \, , 
\end{equation}
and the corresponding resistance $R_{xx}$ reads
\begin{equation}
R_{xx}=1+\frac{h^2-\rho^2}{k^2 r_h^2}+\mathcal{O}(r_h^{-3})=1+\frac{9}{16\pi^2} \frac{h^2-\rho^2}{k^2}\;T^{-2}+\mathcal{O}(T^{-3})\,.
\end{equation}
It is clear that when $h<\rho$, $R_{xx}$ increases monotonically with $T$ (working under the assumption above that the temperature is the largest scale in the problem), displaying metallic behavior.
On the other hand, it decreases with increasing $T$ when $h>\rho$, displaying insulating behavior. In this regime, the Hall component and the inverse Hall angle become
\begin{equation}
R_{yx}=\frac{9}{8\pi^2}\frac{h\rho}{k^2}\;T^{-2}+\mathcal{O}(T^{-3})\,,\quad
\cot\Theta=\frac{8\pi^2}{9}\frac{k^2}{h\rho}\;T^2+\mathcal{O}(T^0)\,.
\end{equation}

At this point one would like to restore all theory parameters~\eqref{adstheory}. The general results in the high temperature regime are given by
\begin{equation}
\begin{split}
R_{xx}&=\frac{1}{z_1}+\frac{V_0^2}{64\pi^2}\frac{h^2 z_1^2-\rho^2}{k^2 z_1^2}\; T^{-2}+\mathcal{O}(T^{-3})\,,\\
R_{yx}&=\frac{V_0^2}{32\pi^2}\frac{h\rho}{k^2 z_1}\;T^{-2}+\mathcal{O}(T^{-3})\,,\quad\quad\quad T\gg (k,\sqrt{\rho},\sqrt{h})\,,\\
 \cot\Theta&=\frac{32\pi^2}{V_0^2}\frac{k^2}{h\rho}\;T^2+\mathcal{O}(T^0)\,.
\end{split}
\end{equation}
The threshold value for the magnetic field is therefore given by $h/\rho=1/z_1$.
Finally, note that in order to generate more arbitrary scalings, one needs to allow for more complicated background geometries, in which the neutral scalar $\phi$ should be dynamical. We turn to this question next.

\section{Non-relativistic Scaling Geometries}
\label{SectionNonRelativistic}

We are now going to examine the behavior of the resistivities for geometries supported by a non-trivial scalar field profile, and which exhibit non-relativistic scalings. We choose the couplings to have the simple exponential form
\begin{equation}
\label{expcouplings}
Z_1(\phi) = z_1 e^{\gamma\, \phi} \, , \quad Z_2(\phi) =  e^{\delta\, \phi}\, ,  \quad V(\phi) = -V_0 \, e^{\eta\, \phi}\, , \quad  Y(\phi) = e^{\alpha\, \phi} \, ,
\end{equation}
in order to look for \emph{exact} scaling solutions, loosely motivated by top-down realizations~\cite{Kiritsis:2016cpm,Charmousis:2010zz}.
However, one should keep in mind that we will assume that such non-relativistic solutions describe the IR of the geometry, and approach AdS in the UV, so that one can adopt the standard AdS/CFT dictionary. To this end,  the scalar potential of~\eqref{expcouplings} should be appropriately modified, to ensure that the scalar $\phi$ can indeed settle to a constant at the boundary.
That this can be done is by now well known, and has been shown explicitly in a variety of cases in the literature. Thus, here we will simply adopt~\ref{expcouplings} and focus on obtaining exact scaling backgrounds.
We focus mostly on cases with no magnetic field, but also include a simple background solution for which $h$ is non-zero.
Scaling solutions for the Einstein-DBI-dilaton system were also studied first in~\cite{Pal:2012zn} and later in~\cite{Tarrio:2013tta}. 
However the models studied in those papers did not include axions, 
and therefore did not incorporate any mechanism for dissipating momentum, 
resulting in an infinite DC conductivity.

\subsection{Hyperscaling-violating solutions without magnetic field}
We are going to parametrize the geometry as in~\eqref{fullansatz},  and look for black brane solutions of the form
\begin{eqnarray}
\label{hvbrane}
B(r) &=& \frac{L^2 r^{\theta-2}}{f(r)} \, , \quad C(r) = r^{\theta-2} \, , \quad  D(r) = r^{\theta-2z} f(r) \, , \\
A&=&A_{t}(r)\,dt\, , \quad \phi(r) = \kappa \ln r \, ,  \quad \psi^1 = k\, x\,,\quad \psi^2=k\,y \, ,
\end{eqnarray}
where the parameters $z$ and $\theta$ are, respectively, the Lifshitz and hyperscaling violating exponents.
In this ansatz we have turned off the magnetic field, $h=0$, for simplicity.  
Note that when the blackening function is trivial, $f(r)=1$, one recovers the standard hyperscaling-violating geometries 
\begin{equation}\label{extreme}
ds^2=r^\theta\left(-\frac{dt^2}{r^{2z}}+\frac{L^2 dr^2}{r^2}+\frac{dx^2+dy^2}{r^2}\right) ,
\end{equation}
which represent the extremal limit of (\ref{hvbrane}) and can be thought of as generalized quantum critical geometries.
Examining Einstein's equations, we immediately find from~\eqref{eqmetric2} that $\kappa$ must obey
\begin{equation}
\label{kappasquared}
\kappa^2 = (\theta-2)(\theta-2z+2) \, ,
\end{equation}
while from the gauge field equation~\eqref{eqA} we find the derivative of $A_t$,
\begin{equation}
\label{Ader}
A_t^\prime = \frac{\rho L \,r^{\theta-z-\delta\kappa-1}}{\sqrt{\rho^2+ z_1^2  r^{2 [\theta-2+(\gamma + \delta)\kappa] }}} \, .
\end{equation}
In order to obtain exact solutions for this system we will make some assumptions on the parameters of the model.
First, notice that the gauge field expression (\ref{Ader}) simplifies drastically when we set
\begin{equation}
\label{specialtheta}
\theta= 2-\kappa(\gamma + \delta) \, ,
\end{equation}
in which case the gauge field obeys the much simpler condition
\begin{equation}
A_t^\prime =  \frac{ \rho L \,r^{\theta-z-\delta\kappa-1}}{\sqrt{\rho^2 + z_1^2 }} \,.
\end{equation}
Combining (\ref{kappasquared}) and (\ref{specialtheta}) in this case yields the following relation between $z$ and $\theta$,  
\begin{equation}
z = 1 + \frac{\theta}{2}+\frac{2-\theta}{2(\gamma+\delta)^2} \, .
\end{equation}
We are most interested in the case in which the stress tensor terms in the field equations originating from the axions appear at the same order in powers of the radial coordinate as terms coming from the metric, neutral scalar and U(1) gauge fields. This motivates us to take
\begin{equation}
\delta=-\alpha = \frac{2}{\kappa}\,, \quad \eta =\gamma \,.
\end{equation}
Finally, using~\eqref{eqphi},~\eqref{eqmetric1} and~\eqref{eqmetric3}, we find an analytic solution for $f(r)$
\begin{equation}\label{blackening}
f(r)=1-\left(\frac{r}{r_h}\right)^{2+z-\theta}\,,
\end{equation}
where $r_h$ is the location of the horizon, and
\begin{eqnarray}
L^2&=&\frac{2}{\delta^2}\frac{(\gamma+3\delta)(\gamma+\delta)+1}{V_0+z_1-k^2-\sqrt{\rho^2+z_1^2}}\,,\\
k^2&=&\frac{\gamma^2+\gamma\delta-1}{\delta^2+\gamma\delta+1}\left(\frac{z_1^2}{\sqrt{\rho^2+z_1^2}}-z_1-V_0\right)-\frac{\rho^2}{\sqrt{\rho^2+z_1^2}}\,.
\end{eqnarray}
We have demanded that the extremal limit is given by~\eqref{extreme}.

Summarizing our results, in the case of vanishing magnetic field we have obtained the following quantum critical geometry, supported by a running scalar, 
\begin{equation}\label{solutiondbi}
\begin{split}
ds^2 & = r^{\theta} \left[-\frac{f(r)}{r^{2z}}dt^2+\frac{L^2}{r^2 f(r)}dr^2+\frac{dx^2+dy^2}{r^2}\right]\,,  \\
\phi & =  \kappa \ln(r),\quad \psi^1=k\,x,\quad \psi^2=k\,y\,,
\end{split}
\end{equation}
with
\begin{equation}
\begin{split}
f(r)&=1-\left(\frac{r}{r_h}\right)^{2+z-\theta},\quad z=\frac{1-\gamma^2+\delta^2}{\delta(\gamma+\delta)},\quad \theta=-\frac{2\gamma}{\delta},\quad \kappa=\frac{2}{\delta}\,,\\
L^2&=\frac{2}{\delta^2}\frac{(\gamma+3\delta)(\gamma+\delta)+1}{V_0+z_1-k^2-\sqrt{\rho^2+z_1^2}}=\frac{(\theta-2)(\theta-z-2)}{V_0+z_1-k^2-\sqrt{\rho^2+z_1^2}}\,,\\
k^2&=\frac{\gamma^2+\gamma\delta-1}{\delta^2+\gamma\delta+1}\left(\frac{z_1^2}{\sqrt{\rho^2+z_1^2}}-z_1-V_0\right)-\frac{\rho^2}{\sqrt{\rho^2+z_1^2}}\,,\\
      &=\frac{2(z-1)}{(2z-\theta)}\left(V_0+z_1-\frac{z_1^2}{\sqrt{\rho^2+z_1^2}}\right)-\frac{\rho^2}{\sqrt{\rho^2+z_1^2}}\,,\\
\alpha&=-\delta,\quad \eta=\gamma, \quad A_t=\frac{L \rho}{(\theta-z-2)\sqrt{\rho^2+z_1^2}}\, r^{\theta-z-2} \, .
\end{split}
\end{equation}
We can also invert the expressions for $z$ and $\theta$ to obtain
\begin{equation}
\gamma=\pm\frac{\theta}{\sqrt{(\theta-2)(\theta-2z+2)}}\,,\quad \delta=\mp\frac{2}{\sqrt{(\theta-2)(\theta-2z+2)}}\,.
\end{equation}
The temperature associated with these solutions has the simple expression
\begin{equation}
T=\frac{|z+2-\theta|}{4\pi L} \, r_h^{-z} \, ,
\end{equation}
and the thermal entropy is therefore 
\be
\label{entropyscaling}
s \sim r_h^{\theta-2}\sim T^{\frac{2-\theta}{z}} \, .
\ee

There are a number of conditions one should impose on these solutions to ensure that they are well-defined and supported by a matter sector that is physical.  Such conditions will lead to constraints on the allowed range of $\{z,\theta\}$, and therefore on the range of theory parameters $\gamma$ and $\delta$.
First, in order for the solution to be real one should demand\,\footnote{Notice that when $z=1$ we have $k^2<0$. Thus, the relativistic case $z=1$ is not allowed when the axions are present in the theory, as it leads to unphysical conditions on the parameters. Moreover, in order to have $k^2>0$ the quantity $V_0+z_1$ should be positive and sufficiently large.}
\begin{equation}\label{real}
k^2>0,\quad L^2>0,\quad (\theta-2)(\theta-2z+2)>0\,.
\end{equation}
Next, the Null Energy Condition (NEC) should be satisfied, \emph{i.e.}
\begin{equation}\label{nec}
T_{\mu\nu}N^\mu N^\nu\geqslant 0  \, ,
\end{equation}
for any null vector $N^\mu N_\mu=0$.  
For the geometry~\eqref{solutiondbi}, the two independent null vectors can be chosen as
\begin{eqnarray}
N^t  =  \frac{1}{\sqrt{f}} r^{z-\theta/2},\quad N^r=\frac{\sqrt{f}}{L} r^{1-\theta/2}\sin\tau,\quad N^x=r^{1-\theta/2}\cos\tau \, ,
\end{eqnarray}
with $\tau=0\; \text{or} \;\pi/2$. The NEC constraints on the scaling exponents are then 
\begin{equation}\label{necconstraint}
(\theta-2)(\theta-2 z+2) \geqslant 0, \quad (z-1) (2+z-\theta)\geqslant 0\,.
\end{equation}
We also note that, in order for the IR region to be defined unambiguously, we want the $(t, x, y)$ components of the (extremal) metric to scale in the same way with $r$.
From the form of the metric in~\eqref{solutiondbi}, this condition can be seen to give
\begin{equation}
\label{IRunamb}
(\theta-2)(\theta-2 z)>0\,.
\end{equation}
The IR is then located where the $(t, x, y)$ metric components vanish:
\begin{equation}\label{goodir}
\begin{split}
& \theta-2>0 \quad \text{and}  \quad  \theta-2z >0  \quad   \Rightarrow \quad  \text{IR at } r=0\,, \\
& \theta-2>0 \quad \text{and}  \quad  \theta-2z >0  \quad   \Rightarrow \quad  \text{IR at } r=\infty \,.
\end{split}
\end{equation}
Finally, to ensure thermodynamic stability we would like the geometry to have positive specific heat\footnote{This condition is not quite necessary and will not change our results by much. For the case with negative specific heat, the extremal geometry still takes the form~\eqref{extreme}, but is obtained by taking $T\rightarrow \infty$. One could obtain a gapped spectrum for the AC conductivity, for example, by incorporating the linear perturbation analysis~\cite{Charmousis:2010zz,Kiritsis:2015oxa}. }. From (\ref{entropyscaling}) we see that this implies
\begin{equation}
z \, (2-\theta)  >0 \, .
\end{equation}

\begin{figure}[ht!]
\begin{center}
\includegraphics[width=.65\textwidth]{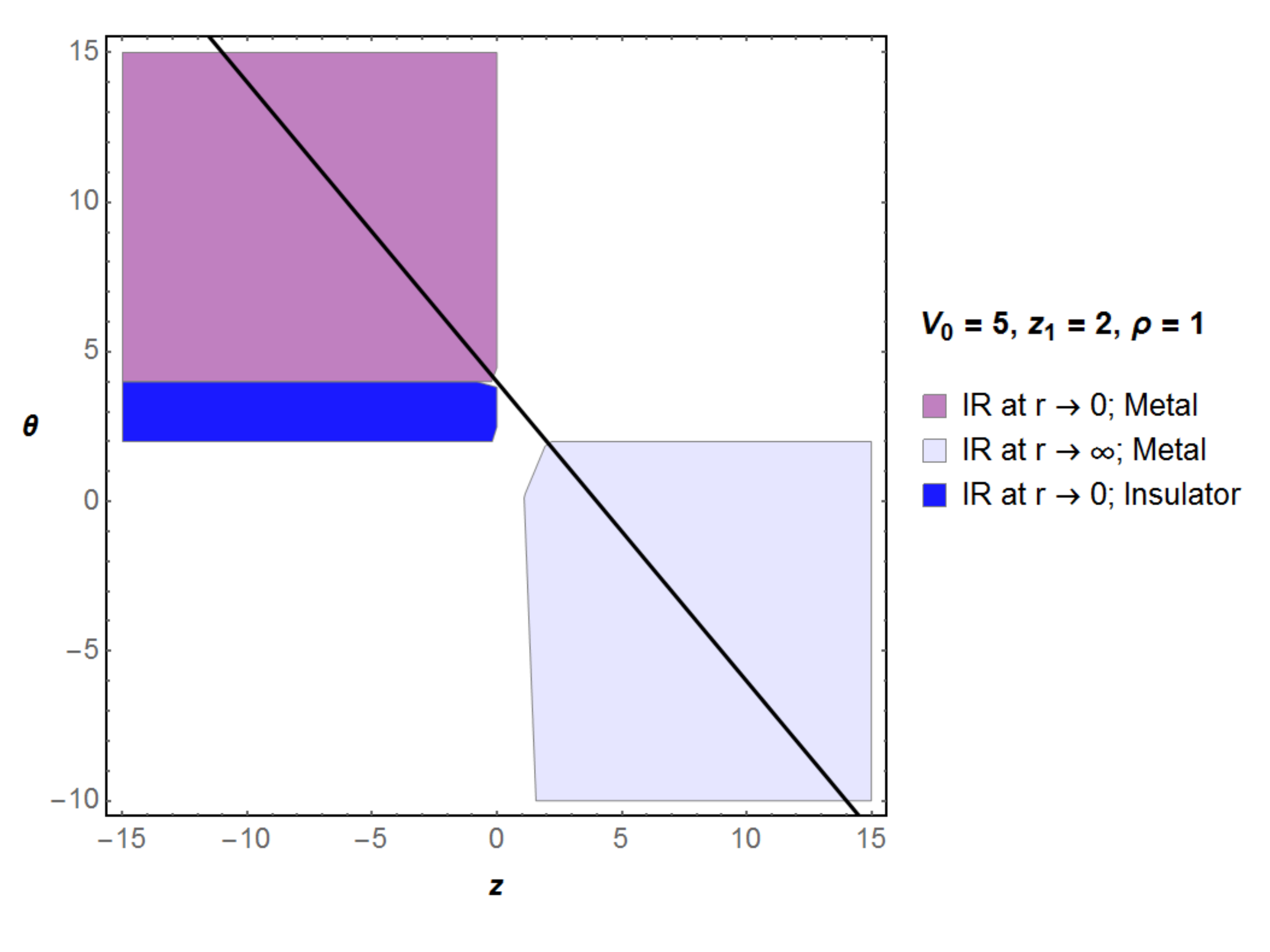}
\caption{The shaded areas denote the allowed ranges of $z$ and $\theta$ after taking into account all constraints on the theory parameter space. 
This case corresponds to $V_0=5, z_1=2, \rho=1$. The straight line $\theta = 4 - z$ corresponds to a resistivity linear in temperature, which for these parameters is allowed in much of the phase space.}
\label{fig:mrdcrhoa}
\end{center}
\end{figure}

\begin{figure}[ht!]
\begin{center}
\includegraphics[width=.65\textwidth]{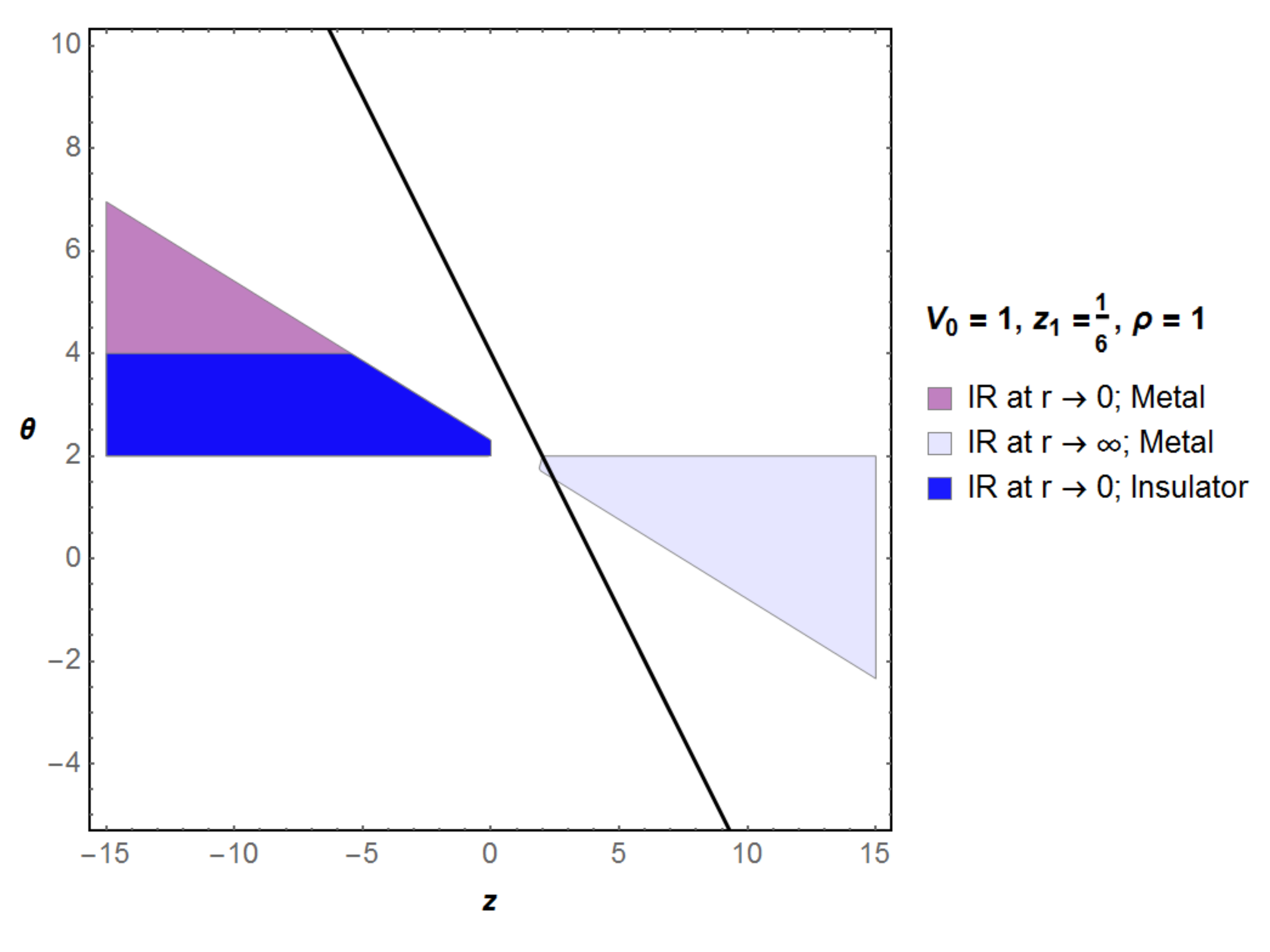}
\caption{The shaded areas denote the allowed ranges of $z$ and $\theta$ after taking into account all constraints on the theory parameter space. 
This case corresponds to $V_0=1, z_1=1/6, \rho=1$. 
The straight line $\theta = 4 - z$ corresponds to a resistivity linear in temperature. }
\label{fig:mrdcrhob}
\end{center}
\end{figure}

Figures~\ref{fig:mrdcrhoa} and~\ref{fig:mrdcrhob} show the allowed ranges of $z$ and $\theta$ which satisfy all of the constraints above, for two different choices of Lagrangian parameters $V_0$ and $z_1$. The charge density has been scaled to $\rho=1$ in both plots. Notice that as $V_0+z_1$ becomes smaller, the allowed parameter space decreases (disappearing completely when $V_0 + z_1$ is negative). 
The figures also indicate whether the UV is located at $r=0$ or $r=\infty$, for a particular region of parameter space.

Armed with these geometries, we can now inspect the behavior of the conductivity.
Substituting the solution into~\eqref{dcnoh}, the expression for $\sigma_{DC}$ in the absence of magnetic field, we find that all the terms scale in the same way with temperature, yielding the simple expression 
\begin{equation}
\sigma_{DC}\sim r_h^{4-\theta}\sim T^{\frac{\theta-4}{z}}.
\end{equation}
Here we see clearly the system behaving as a metal or as an insulator, according to (\ref{criterion}), depending on the sign of $z$ and the range of $\theta$.
Figures~\ref{fig:mrdcrhoa} and~\ref{fig:mrdcrhob} also display the parameter ranges associated with metallic or insulating behavior. 
Note that in this model there is no obstruction to obtaining a linear resistivity\footnote{See {\em e.g.}~\cite{Hartnoll:2009ns,Lee:2010ii,Kim:2010zq,Pal:2010sx,Karch:2014mba}  for the study of holographic strange metals in the probe DBI approximation.}. Indeed, requiring the latter singles out a line in parameter space,
\begin{equation}
\label{linearres}
\theta+z=4 \quad \Rightarrow \quad R_{DC}=\frac{1}{\sigma_{DC}} \sim T \, ,
\end{equation}
which corresponds to taking $\delta = -\gamma \pm \frac{1}{\sqrt{3}}$. 
The linear resistivity case is indicated by the solid line in the figures. Notice that it is allowed in most of the 
parameter space, provided that $V_0+z_1$ is sufficiently large and positive. 

\subsection{Dyonic solutions and negative magnetoresistance }
\begin{figure}[ht!]
\begin{center}
\includegraphics[width=.45\textwidth]{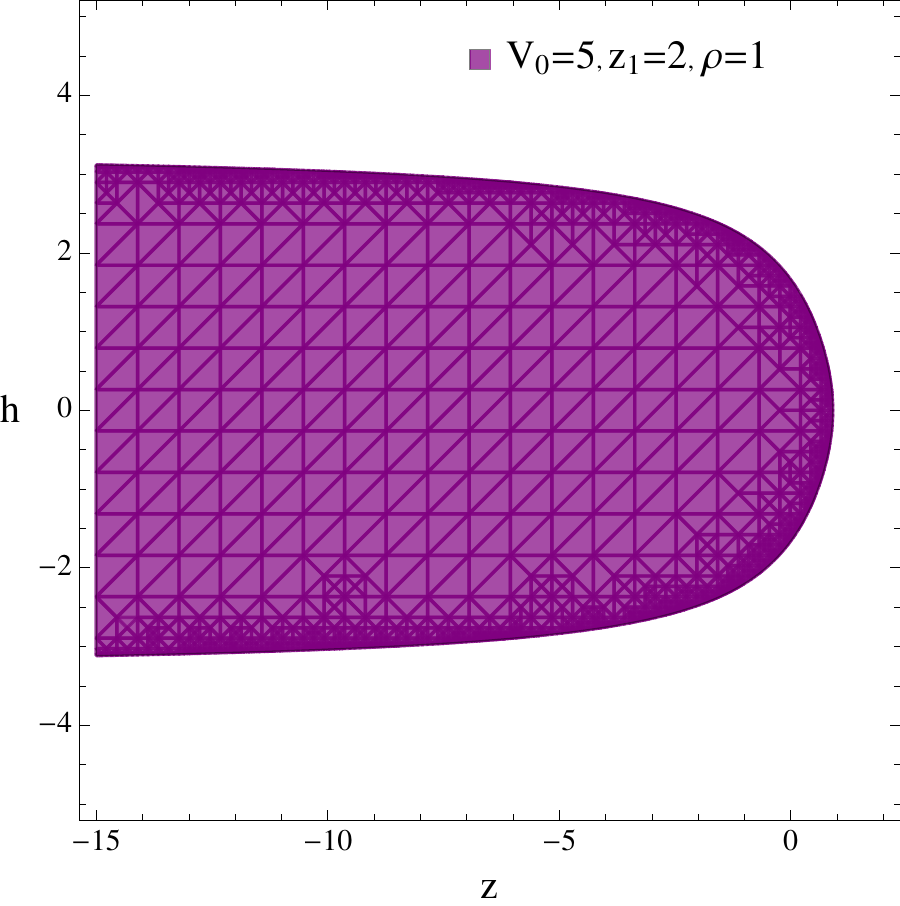}\quad\quad
\includegraphics[width=.45\textwidth]{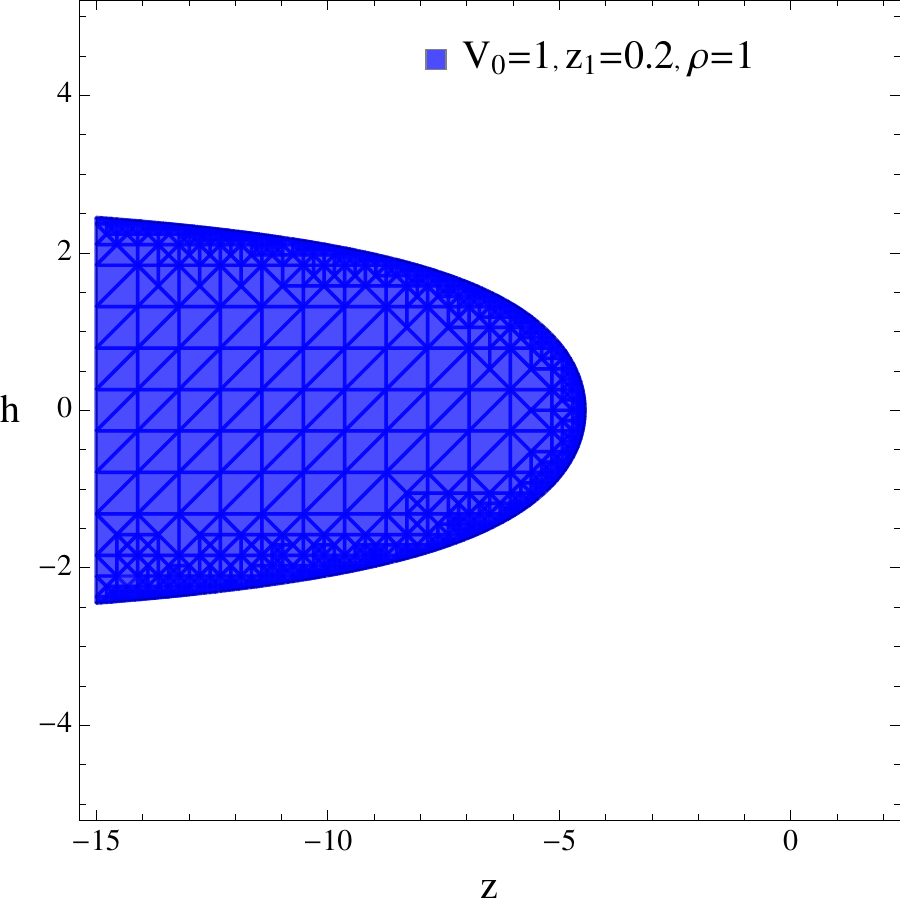}
\caption{The shaded area denotes the allowed ranges of $z$ and $h$ after taking into account the NEC, 
condition (\ref{IRunamb}) and requiring all theory parameters to be real. }
\label{fig:magn}
\end{center}
\end{figure}

For the solutions we have just examined the background magnetic field vanishes.  
It is much more difficult to obtain analytic dyonic solutions, for which both the electric charge density and the magnetic charge are non-trivial. 
Here we show a simple family of exact solutions we obtained after turning on $h$:
\begin{eqnarray}\label{caseh}
f(r)&=&1-\left(\frac{r}{r_h}\right)^{z-2},\quad z=3-\frac{4}{\gamma^2},\quad \theta=4,\quad \kappa=-\frac{4}{\gamma} \, , \nn \\
L^2&=&\frac{2}{\gamma^2}\frac{4-\gamma^2}{V_0+z_1-k^2-\sqrt{\rho^2+z_1^2(1+h^2)}}=\frac{2(2-z)}{V_0+z_1-k^2-\sqrt{\rho^2+z_1^2(1+h^2)}}\, , \nn \\
k^2&=&\frac{(z-1)}{(z-2)}\left[V_0+z_1-\frac{z_1^2}{(z-1)}\frac{(z-1)+h^2(z-2)}{\sqrt{\rho^2+z_1^2(1+h^2)}}\right]-\frac{\rho^2}{\sqrt{\rho^2+z_1^2(1+h^2)}}\, , \nn \\
\alpha&=&-\delta=\frac{\gamma}{2},\quad \eta=\gamma, \quad A_t=\frac{L \rho}{(2-z)\sqrt{\rho^2+z_1^2(1+h^2)}}\, r^{2-z}.
\end{eqnarray}
Note that the hyperscaling violating exponent is fixed in these geometries, \emph{i.e.} $\theta=4$ (and as a result, the IR is always located at $r=0$).
In addition to the NEC, the parameters must be chosen in such a way to ensure that both $k^2$ and $L^2$ are positive. Moreover, condition (\ref{IRunamb}) must hold.
The parameter space for the Lifshitz exponent $z$ and the magnetic field $h$ allowed by these constraints is shown in Figure~\ref{fig:magn}, for  different choices of $V_0$ and $z_1$  (we did not require the specific heat to be positive, which in this case would only change the plots slightly).
As in the previous solution, the phase space becomes smaller as the quantity $V_0+z_1$ decreases.

\begin{figure}[ht!]
\begin{center}
\includegraphics[width=.45\textwidth]{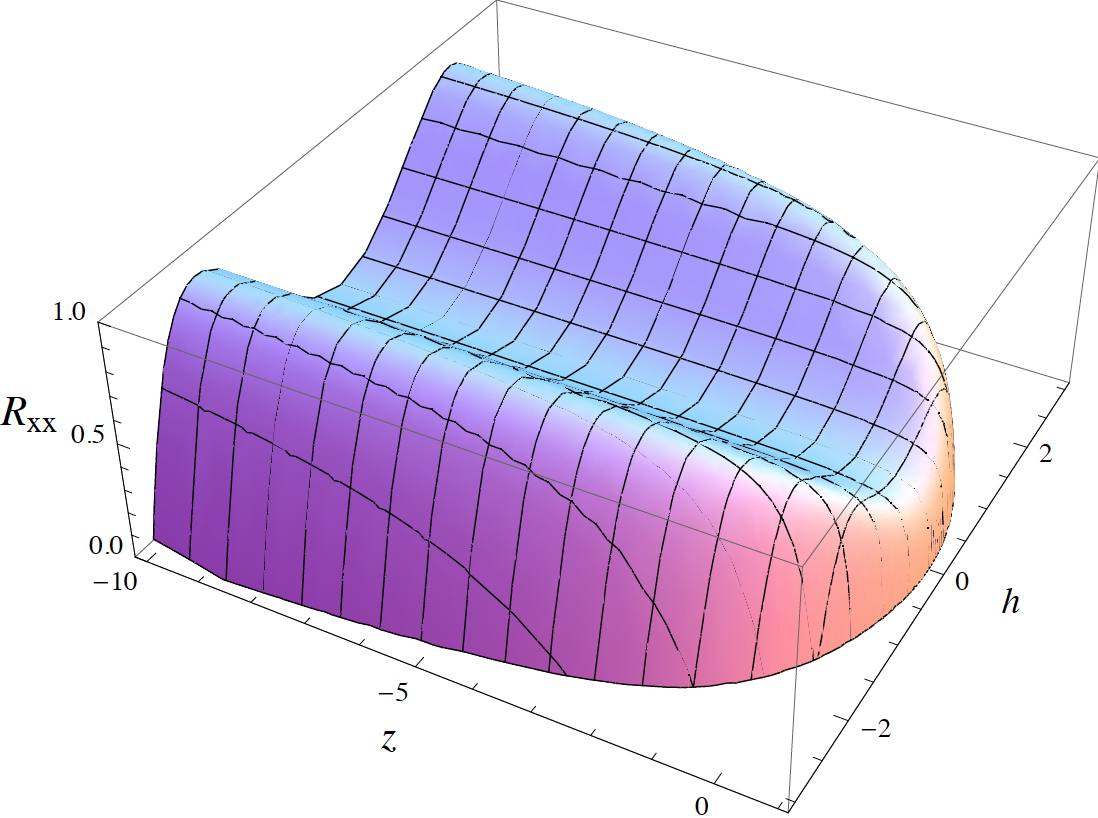}\quad
\includegraphics[width=.45\textwidth]{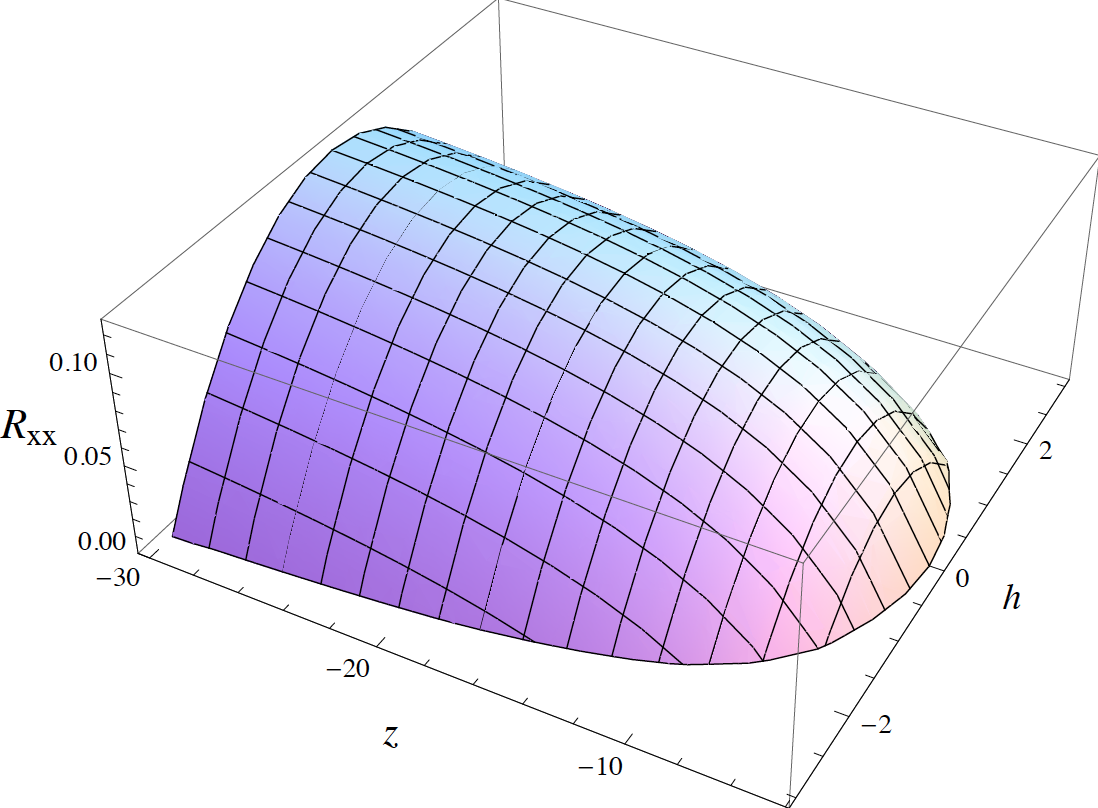}
\caption{The in-plane resistance $R_{xx}$ versus $h$ for various values of $z$. 
It is an even function of $h$. Left panel: $V_0=5, z_1=2$. Right panel: $V_0=1, z_1=0.2$. The charge density has been fixed to $\rho=1$.}
\label{fig:dcxxh}
\end{center}
\end{figure}

Substituting the solution~\eqref{caseh} into~\eqref{DC} we obtain the conductivity matrix, which now depends only on the magnetic field, and not on the temperature. $R_{xx}$ is an even function as a function of $h$, while the Hall part $R_{xy}$ is an odd function. The resistance $R_{xx}$ as a function of $h$ for different values of $z$ is presented in Figure~\ref{fig:dcxxh}. We find that the physical constraint shown in Figure~\ref{fig:magn} ensures that $R_{xx}$ is not negative, as required for a well defined theory. The dual system falls into a particular quantum critical regime where the transport property is determined solely by the magnetic field, independent of the temperature. Depending on the choice of theory parameters $V_0$ and $z_1$,  $R_{xx}$ versus $|h|$ can have a non-monotonic (left panel) or monotonic (right panel) behavior.

Next, let's consider the behavior of the magnetoresistance supported by these solutions. Recall that the 
standard definition of magnetoresistance is given by
\begin{equation}
MR=\frac{R_{xx}(h)-R_{xx}(h=0)}{R_{xx}(h=0)}\,,
\end{equation}
describing the tendency of a material to change the value of its electrical resistance in an externally-applied magnetic field. In particular, a negative magnetoresistance has been observed in many materials, see {\em e.g.}~\cite{Woods:1964,Negisgi:1997,zhou:2011,Huang:2015,Caizhen:2015,Lihui:2015,Culo:2017,Jiang:2017}.
Interestingly, one finds from Figure~\ref{fig:dcxxh} that our system also exhibits negative magnetoresistance\footnote{It would be interesting to generalize our discussion to higher-dimensional theories, in which we would have a longitudinal channel along the magnetic field and a transverse channel perpendicular to it. In certain cases one can obtain a negative longitudinal magnetoresistance
(see \emph{e.g.} \cite{Goswami:2015uxa,Baumgartner:2017kme}).}.
The left panel of Figure~\ref{fig:dcxxh} shows a positive value of MR in the regime of small magnetic field. However, it is easy to see that in the right panel the magnetoresistance is negative, in all of the allowed parameter range. 
We emphasize that such negative MR would not be seen in the probe DBI limit.
Indeed, as one can see from~\eqref{rhoxx}, which is the probe approximation result, $R_{xx}(h)$ increases monotonically with the magnetic field, resulting in a positive MR value.  As a side note, the case with a trivial dilatonic scalar $\phi$ examined in Section~\ref{metalinsulator} 
also has positive magnetoresistance, as can be seen in Figures~\ref{fig:mrdc} and~\ref{fig:mrdcl}. A negative MR is obtained in the quantum critical region characterized by $(\theta=4,z)$. However, notice that for these solutions $k$ is not independent of $h$. In particular, this means the
requirement that the system remains in a given quantum critical regime (described by a fixed $\theta,z$) imposes a non-trivial relation between $k$ and $h$.

In closing this discussion, it is worth noting that we have identified interesting features even in this very simple dyonic setup, using the geometry~\eqref{caseh}.
More generic solutions with $h\neq0$ and broader ranges of $z$ and $\theta$ can, in principle, be obtained, but are significantly more complicated and have therefore been omitted. We expect them to have non-trivial magnetotransport properties and to lead to the same kinds of metal-insulator transitions we discussed in Section \ref{metalinsulator}. Since $h$ is an adjustable parameter, it is also interesting to see if one could obtain a quantum phase transition by tuning the magnetic field.
We leave the analysis of these cases to future work.


\subsection{Solutions with $AdS_2$ geometry}
In the discussion above we have focused on scaling solutions driven by a runaway scalar deep inside the bulk, $\phi_{IR}\rightarrow \infty$. These kinds of solutions emerge when we allow the coupling functions to have the simple exponential form~\eqref{expcouplings}, loosely motivated by top-down string theory realizations.
As we already mentioned, even though these scaling solutions are exact, we are interested 
in the case in which they describe only the near horizon region of the spacetime at low temperatures.

However, our theory with the simple couplings~\eqref{expcouplings} allows for much richer solutions, including some for which in the IR the scalar approaches a constant at extremality. A simple example is given by the following $h=0$ geometry.
\begin{equation}
\begin{split}
f(r)=&1-\left(\frac{r}{r_h}\right)^{2+z-\theta}+\frac{V_0 L^2}{(\theta-2)(\theta+z-4)}r^{2z-2}\left[1-\left(\frac{r}{r_h}\right)^{4-z-\theta}\right]\,,\\
z =& \frac{1-\gamma^2+\delta^2}{\delta(\gamma + \delta)} \, , \quad \theta = -\frac{2\gamma}{\delta} \, , \quad \kappa = \frac{2}{\delta} \, ,\\
k^2=&\frac{2(z-1)}{(2z-\theta)}\left(z_1-\frac{z_1^2}{\sqrt{\rho^2+z_1^2}}\right)-\frac{\rho^2}{\sqrt{\rho^2+z_1^2}}\,,\\
L^2 =&(\theta-2z)(\theta-z-2)\frac{\sqrt{\rho^2+z_1^2}(z_1+\sqrt{\rho^2+z_1^2})}{z_1\rho^2}\,,  \\
\alpha =& -\delta\,, \quad \eta = \frac{1}{\gamma + \delta}\,,\quad A_t(r) =\frac{L \rho}{(\theta-z-2)\sqrt{\rho^2+z_1^2}}r^{\theta-z-2}\,, \\
\gamma=&\pm\frac{\theta}{\sqrt{(\theta-2)(\theta-2z+2)}},\quad \delta=\mp\frac{2}{\sqrt{(\theta-2)(\theta-2z+2)}}\,.
\end{split}
\end{equation}
whose blackening factor is much more involved than that (\ref{blackening}) appearing in the solutions we discussed above.
As a consequence, the temperature associated with these geometries has a more complicated dependence on $r_h$, and is given by
\begin{equation}\label{temnew}
T=\frac{r_{h}^{1-z}}{4\pi L}|f'(r_h)|=\frac{1}{4\pi L} \left|(z+2-\theta)r_h^{-z}-\frac{V_0 L^2}{(\theta-2)}r_{h}^{z-2}\right| \, .
\end{equation}
Substituting this background into~\eqref{dcnoh}, the expression for $\sigma_{DC}$ in the absence of magnetic field, we obtain 
\begin{equation}
\sigma_{DC}\sim r_h^{4-\theta}.
\end{equation}
Using~\eqref{temnew} one can then convert $r_h$ to temperature, obtaining the DC conductivity as a function of $T$.

%
\begin{figure}[ht!]
\begin{center}
\includegraphics[width=.42\textwidth]{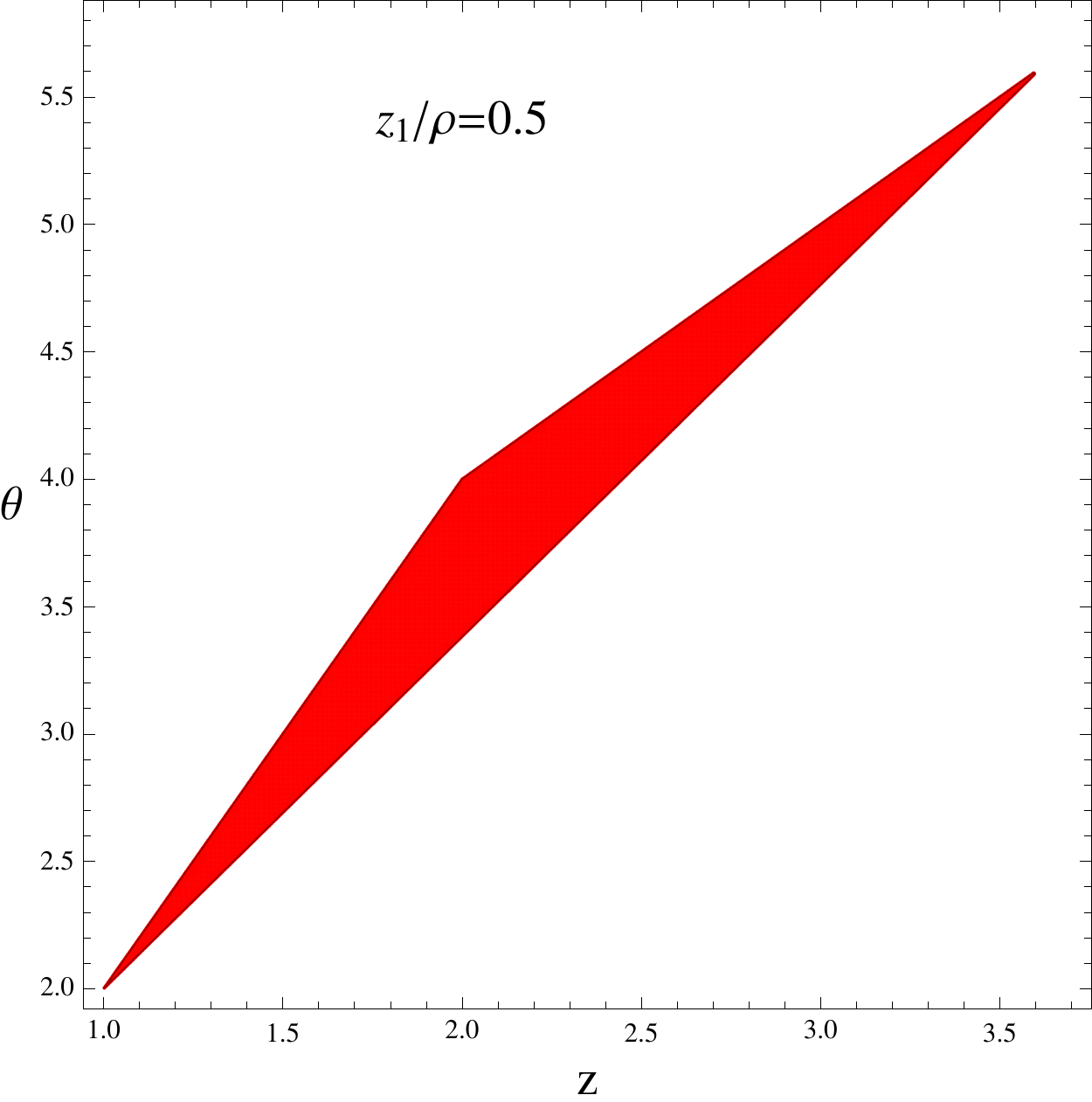}\quad\quad
\includegraphics[width=.41\textwidth]{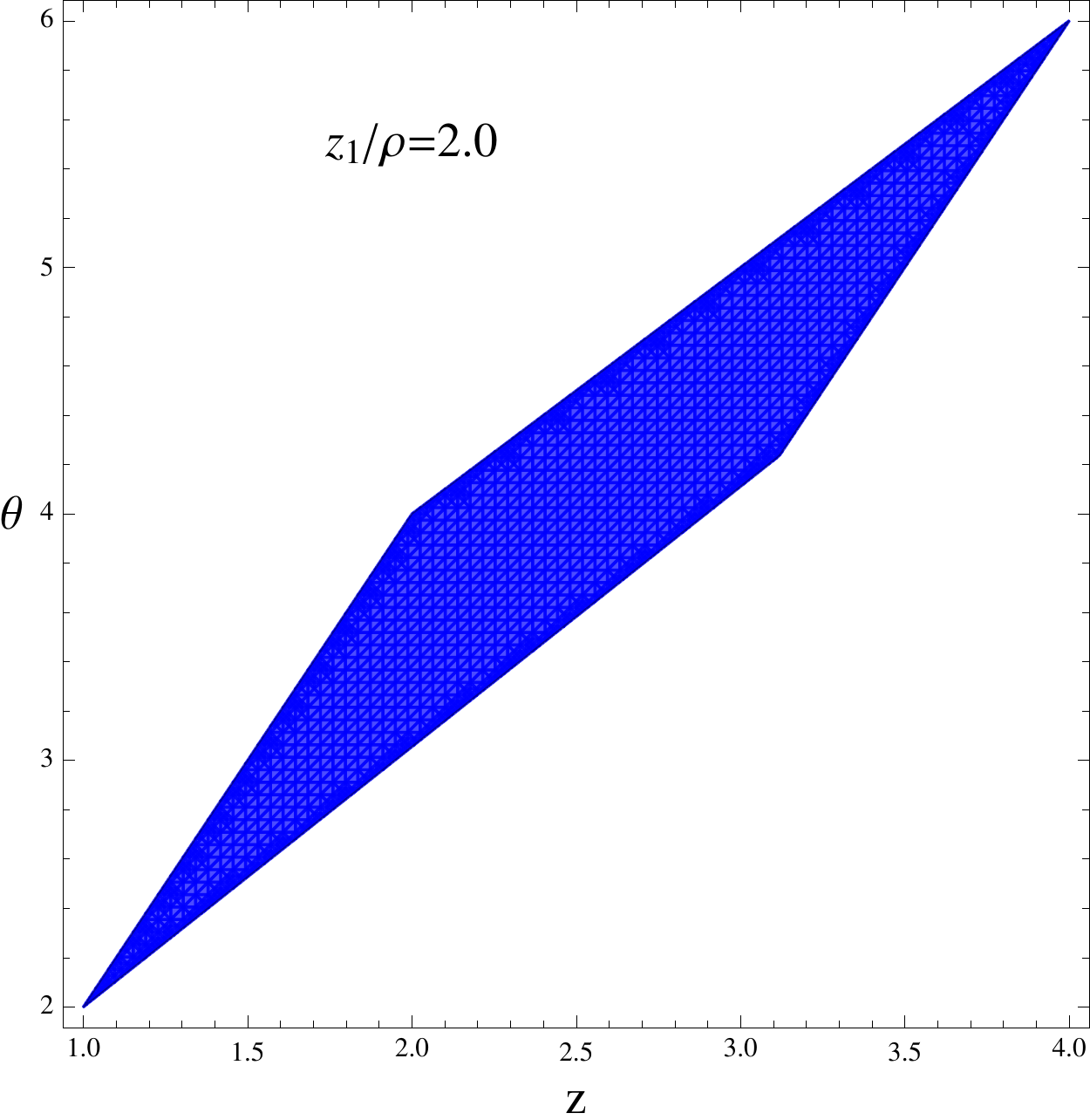}
\caption{The shaded areas denote the allowed ranges of $\{\theta, z\}$ after considering the constraint~\protect\eqref{real}. The range of the parameter space depends on the ratio $z_1/\rho$.}
\label{fig:newz1}
\end{center}
\end{figure}

Note that by letting $V_0=0$ one could naively recover the standard hyperscaling-violating geometry with the blackening factor given by~\eqref{blackening}. 
The choice $V_0=0$, however, is not consistent with the various constraints on the parameter space (NEC and the requirement that $L$ and $k$ are real), as can be checked. Thus, $V_0$ must be taken to be non-vanishing, and the resulting solution is intrinsically different 
from the usual one~\eqref{extreme} with~\eqref{blackening}. 
As a result, some of the constraints, such as~\eqref{goodir}, no longer apply. 
The two restrictions one can impose are the NEC~\eqref{nec} and the reality of all metric/scalar/gauge field coefficients. The latter demands
\begin{equation}\label{real}
L^2\sim (\theta-2z)(\theta-z-2)>0,\quad k^2>0,\quad (\theta-2)(\theta-2z+2)>0\,.
\end{equation}
The NEC gives the same condition as~\eqref{necconstraint}. Moreover, one can show that~\eqref{necconstraint} is already implied  by the requirement that the solution be real.
The resulting parameter range for the exponents $\{\theta,z\}$ depends on the ratio $z_1/\rho$. Toy examples are shown in Figure~\ref{fig:newz1}.  In particular, one finds from Figure~\ref{fig:newz1} that $(\theta-2)>0$ and $(z+2-\theta)>0$, conditions which can be shown to be valid in general by considering~\eqref{real}.

Therefore, according to~\eqref{temnew}, the extremal limit $T\rightarrow 0$ is obtained at a finite value of the horizon radius, 
\begin{equation}
\hat{r}_h=\left(\frac{(\theta-2)(z+2-\theta)}{V_0 L^2}\right)^{1/(2z-2)}\,. 
\end{equation}
Indeed, the extremal near-horizon geometry for these solutions contains an $AdS_2$ factor, supported by a constant scalar 
$\phi_{IR}=\phi(\hat{r}_h)$, and is therefore associated with a finite entropy density.  
The DC conductivity $\sigma_{DC}$ is also finite as $T\rightarrow 0$, unlike the hyperscaling-violating solution~\eqref{solutiondbi}. 
At finite temperature, for certain choices of $\{\theta, z\}$ -- or equivalently $\{\gamma, \delta\}$ -- the resistivity will decrease as $T$ is lowered, showing metallic behavior  
according to the criterion~\eqref{criterion}.
However, it can also increase as the temperature decreases, which is reminiscent of an insulator. 
We are going to postpone a more thorough study of the transport properties associated with these solutions to future work. For now it suffices to say that they provide a concrete example 
of the richness of the near horizon backgrounds allowed in these constructions, even assuming the simple choice of scalar couplings (\ref{expcouplings}).


\addcontentsline{toc}{section}{Acknowledgments}
\acknowledgments\label{ACKNOWL}

We would like to thank Elias Kiritsis for useful comments on the draft.
The work of S.C. and A.H. is supported in part by the National Science Foundation grant PHY-1620169.

\appendix
\renewcommand{\theequation}{\thesection.\arabic{equation}}
\addcontentsline{toc}{section}{Appendix}
\section*{Appendices}

\section{Born-Infeld Theory}
\label{Appendix}

Another well-known non-linear generalization of Maxwell's electromagnetism is Born-Infeld theory, whose structure is similar to that of DBI theory, but somewhat simpler.
 From a phenomenological point of view, it might also be interesting to consider magnetotransport in this case.
In particular, although the conductive behavior shares the same overall features as that of DBI, its dependence on the couplings of the theory is simpler.
As a consequence, the identification of scaling regimes might be easier in the context of Born-Infeld interactions. 
With future applications in mind, here we present the general formula for magnetotransport for Born-Infeld theory. 

The action we consider reads
\begin{equation}\label{app:action}
S=\int d^{4}x\sqrt{-g}\left[\mathcal{R}-{1\over 2}(\partial\phi)^2-V(\phi)-\frac{Y(\phi)}{2}\sum_{I=1}^2 (\partial \psi^I)^2-\frac{Z_a(\phi)}{2}(\sqrt{1+Z_b(\phi)\,F^2}-1)  \right]\,,
\end{equation}
with the three couplings $(Z_a, Z_b, Y)$ chosen to be functions of the neutral scalar $\phi$. 
The last term in (\ref{app:action}) is precisely the non-linear interaction known as the Born-Infeld term.
We consider the same homogeneous and isotropic background as in~\eqref{fullansatz}, and obtain the charge density
\begin{equation}\label{app:charge}
\rho=\frac{\sqrt{-g}Z_a(\phi)Z_b(\phi) F^{t r}}{\sqrt{1+Z_b(\phi)\,F^2}}=\frac{C Z_a Z_b A_t'}{\sqrt{1+Z_b\left(\frac{2 h^2}{C^2}-2 A_t'^2\right)}}\,.
\end{equation}
To calculate the transport coefficients we adopt the same method of Section 3, and find that the
conductivity matrix is given by
\begin{equation}\label{app:DC}
\begin{split}
\sigma_{xx}&=\sigma_{yy} =  \frac{k^2\rho^2 C Y+ k^2 C Y \Sigma(k^2 C Y+h^2\Sigma)}{h^2 \rho^2+(k^2 C Y+h^2 \Sigma)^2}\Big{|}_{r=r_h}\, ,\\
\sigma_{xy}&=-\sigma_{yx}=  \frac{2 k^2 h \rho C Y\Sigma+h \rho(\rho^2+h^2\Sigma^2)}{h^2 \rho^2+(k^2 C Y+h^2 \Sigma)^2}\Big{|}_{r=r_h}\,,
\end{split}
\end{equation}
with the function $\Sigma(r)$ defined to be given by
\begin{equation}
\Sigma(r)=\sqrt{\frac{2 Z_b(\phi)\rho^2+C(r)^2 Z_a(\phi)^2 Z_b(\phi)^2}{C(r)^2+2 h^2 Z_b(\phi)}}\,.
\end{equation}
We see that the conductivity matrix is controlled by the three scalar couplings $Z_a, Z_b, Y$ and by the spatial metric component $C$, all evaluated at the horizon $r=r_h$. They will depend on temperature $T$ (through their dependence on $r_h$), magnetic field $h$ as well as the amount of momentum dispassion $k$ and charge density $\rho$.

The inverse Hall angle is then given by
\begin{equation}
\cot \Theta_H=\frac{\sigma_{xx}}{\sigma_{xy}}=\frac{k^2\rho^2 C Y+ k^2 C Y \Sigma(k^2 C Y+h^2 \Sigma)}{2 k^2 h \rho C Y\Sigma+h \rho(\rho^2+h^2\Sigma^2)}\,,
\end{equation}
and the resistivity matrix, obtained by inversting (\ref{app:DC}), reads
\begin{equation}\label{app:resis}
\begin{split}
R_{xx}&=R_{yy}=\frac{\sigma_{xx}}{\sigma_{xx}^2+\sigma_{xy}^2} =  \frac{k^2 CY(\rho^2+k^2 C Y\Sigma+h^2\Sigma^2)}{h^2 \rho^2\Sigma^2+(\rho^2+k^2 CY\Sigma)^2}\,,\\
R_{xy}&=-R_{yx}=-\frac{\sigma_{xy}}{\sigma_{xx}^2+\sigma_{xy}^2} = \frac{h\rho(\rho^2+2k^2 C Y\Sigma+h^2\Sigma^2)}{h^2 \rho^2\Sigma^2+(\rho^2+k^2 CY\Sigma)^2}\,.
\end{split}
\end{equation}
Since these expressions are quite involved, we would like to restrict our attention to three simple cases:
\begin{itemize}
 \item No momentum dissipation: in the limit $k\rightarrow 0$, the momentum dissipation disappears and the conductivity tensor becomes
\begin{equation}
\sigma_{xx}=\sigma_{yy}=0,\quad \sigma_{xy}=-\sigma_{yx}=\frac{\rho}{h}\, ,
\end{equation}
the same result we obtained for the DBI case studied in the main text.
As before, in this limit the conductivities are independent of the temperature and the details of the theory. 
  \item No magnetic field: after turning off the background magnetic field, we obtain
\begin{equation}
\sigma_{DC}=\sigma_{xx}=\sqrt{Z_a^2 Z_b^2+\frac{2 Z_b \rho^2}{C^2}}+\frac{\rho^2}{k^2 C Y}=\sqrt{Z_a^2 Z_b^2+\frac{32\pi^2 Z_b \rho^2}{s^2}}+\frac{4\pi \rho^2}{k^2 Y s},
\end{equation}
where $s=4\pi \,C(r_h)$ is the entropy density. The dependence on the couplings of the theory is now slightly different from that of the DBI case, as expected from the different structure of the action. 
Note that $\sigma_{xy}=0$ when $h=0$. 
A particular simple case with $Z_{a,b}$ and $Y$ constants and without the neutral scalar $\phi$ was discussed in~\cite{Baggioli:2016oju}, where some features of Mott-like states  were identified.
  \item No charge density: The DC resistivity when $\rho = 0$ reads
\begin{equation}
R_{DC}=R_{xx}=\frac{1}{Z_a Z_b}\sqrt{1+\frac{2 Z_b}{C^2}h^2}+\frac{h^2}{k^2 C Y}=\frac{1}{Z_a Z_b}\sqrt{1+\frac{32\pi^2 Z_b}{s^2}h^2}+\frac{4\pi h^2}{k^2 Y s},
\end{equation}
and the Hall part vanishes, $R_{xy}=0$.
\end{itemize}
The explicit dependence on the two scales $T$ and $h$ can be determined after substituting specific background geometries into the general resistivity expressions. However, finding black hole solutions for the Born-Infeld-Axion theory~\eqref{app:action} is beyond the scope of this paper, and is delegated to future work.



\end{document}